# Deep learning tackles single-cell analysis – A survey of deep learning for scRNA-seq analysis

**Mario Flores[1§], Zhentao Liu[1], Tinghe Zhang[1], Md Musaddaqui Hasib[1], Yu-Chiao Chiu[2], Zhenqing Ye[2,3], Karla Paniagua[1], Sumin Jo[1], Jianqiu Zhang[1], Shou-Jiang Gao[4,6], Yu-Fang Jin[1], Yidong Chen[2,3§], and Yufei Huang[5,6§]**

[1]Department of Electrical and Computer Engineering, the University of Texas at San Antonio, San Antonio, TX 78249, USA

[2]Greehey Children's Cancer Research Institute, University of Texas Health San Antonio, San Antonio, TX 78229, USA

[3]Department of Population Health Sciences, University of Texas Health San Antonio, San Antonio, TX 78229, USA

[4]Department of Microbiology and Molecular Genetics, University of Pittsburgh, Pittsburgh, Pennsylvania, PA 15232, USA

[5]Department of Medicine, School of Medicine, University of Pittsburgh, PA 15232, USA

[6]UPMC Hillman Cancer Center, University of Pittsburgh, PA 15232, USA

[§]**Correspondence** should be addressed to Mario Flores (mario.flores@utsa.edu); Yidong Chen (cheny8@uthscsa.edu); Yufei Huan (yuh119@pitt.edu)

**Running title**: Deep learning for single-cell RNA-seq analysis

**Mario Flores**, Ph.D., is an Assistant Professor in the Department of Electrical and Computer Engineering at the University of Texas at San Antonio, His research focuses on DNA and RNA sequence methods, transcriptomics analysis (including scRNA-Seq), epigenetics, comparative genomics, and deep learning to study mechanisms of gene regulation.

**Zhentao Liu** is a Ph.D. student in the Department of Electrical and Computer Engineering, the University of Texas at San Antonio. His research focuses on deep learning for cancer genomics and drug response prediction.

**Tinghe Zhang** is a Ph.D. student in the Department of Electrical and Computer Engineering, the University of Texas at San Antonio. His research focuses on deep learning for cancer genomics and drug response prediction.

**Md Musaddaqui Hasib** is a Ph.D. student in the Department of Electrical and Computer Engineering, the University of Texas at San Antonio. His research focuses on interpretable deep learning for cancer genomics.

**Zhenqing Ye** Ph.D., is an Assistant Professor in the Department of Population Health Sciences and the director of Computational Biology and Bioinformatics at Greehey Children's Cancer Research Institute at the University of Texas Health San Antonio. His research focuses on computational methods on next-generation sequencing and single-cell RNA-seq data analysis.

**Sumin Jo** is a Ph.D. student in the Department of Electrical and Computer Engineering, the University of Texas at San Antonio. Her research focuses on $m^6A$ mRNA methylation and deep learning for biomedical applications.

**Karla Paniagua** is a Ph.D. student in the Department of Electrical and Computer Engineering, the University of Texas at San Antonio. Her research focuses on applications of deep learning algorithms.

**Yu-Chiao Chiu.** Ph.D., is a Postdoctoral Fellow at the Greehey Children's Cancer Research Institute at the University of Texas Health San Antonio. His postdoctoral research is focused on developing deep learning models for pharmacogenomic studies.

**Jianqiu Zhang,** PhD, is an Associate Professor in the Department of Electrical and Computer Engineering at the University of Texas at San Antonio. Her current research focuses on deep learning for biomedical applications such as $m^6A$ mRNA methylation.

**Shou-Jiang Gao,** Ph.D., is a Professor in UPMC Hillman Cancer Center and Department of Microbiology and Molecular Genetics, University of Pittsburgh. His current research interests include Kaposi's sarcoma-associate herpesvirus (KSHV), AIDS-related malignancies, translational and cancer therapeutics, and systems biology.

**Yu-Fang Jin**, Ph.D., is a Professor in the Department of Electrical and Computer Engineering at the University of Texas at San Antonio. Her research focuses on mathematical modeling of cellular responses in immune systems, data-driven modeling and analysis of macrophage activations, and deep learning applications.

**Yidong Chen**, Ph.D., is a Professor in the Department of Population Health Sciences and the director of Computational Biology and Bioinformatics at Greehey Children's Cancer Research Institute at the University of Texas Health San Antonio. His research interests include bioinformatics methods in next-generation sequencing technologies, integrative genomic data analysis, genetic data visualization and management, and machine learning in translational cancer research

**Yufei Huang**, Ph.D., is a Professor in UPMC Hillman Cancer Center and Department of Medicine, University of Pittsburgh School of Medicine. His current research interests include uncovering the functions of $m^6A$ mRNA methylation, cancer virology, and medical AI & deep learning.

## Abstract

Since its selection as the method of the year in 2013, single-cell technologies have become mature enough to provide answers to complex research questions. With the growth of single-cell profiling technologies, there has also been a significant increase in data collected from single-cell profilings, resulting in computational challenges to process these massive and complicated datasets. To address these challenges, deep learning (DL) is positioning as a competitive alternative for single-cell analyses besides the traditional machine learning approaches. Here we present a processing pipeline of single-cell RNA-seq data, survey a total of 25 DL algorithms and their applicability for a specific step in the processing pipeline. Specifically, we establish a unified mathematical representation of all variational autoencoder, autoencoder, and generative adversarial network models, compare the training strategies and loss functions for these models, and relate the loss functions of these models to specific objectives of the data processing step. Such presentation will allow readers to choose suitable algorithms for their particular objective at each step in the pipeline. We envision that this survey will serve as an important information portal for learning the application of DL for scRNA-seq analysis and inspire innovative use of DL to address a broader range of new challenges in emerging multi-omics and spatial single-cell sequencing.

## Key points:

- Single cell RNA sequencing technology generates a large collection of transcriptomic profiles of up to millions of cells, enabling biological investigation of hidden expression functional structures or cell types, predicting their effects or responses to treatment more precisely, or utilizing subpopulations to address unanswered hypotheses.
- Current deep learning-based analysis approaches for single cell RNA seq data are systematically reviewed in this paper according to the challenge they address and their roles in the analysis pipeline.
- A unified mathematical description of the surveyed DL models is presented and the specific model features were discussed when reviewing each approach.
- A comprehensive summary of the evaluation metrics, comparison algorithms, and datasets by each approach is presented.

## 1. Introduction

Single cell sequencing technology has been a rapidly developing area to study genomics, transcriptomics, proteomics, metabolomics, and cellular interactions at the single cell level for cell-type identification, tissue composition, and reprogramming [1, 2]. Specifically, sequencing of the transcriptome of single cells, or single-cell RNA-sequencing (scRNA-seq), has become the dominant technology in many frontier research areas such as disease progression and drug discovery [3, 4]. One particular area where scRNA-seq has made a tangible impact is cancer, where scRNA-seq is becoming a powerful tool for understanding invasion, intratumor heterogeneity, metastasis, epigenetic alterations, detecting rare cancer stem cells, and therapeutic response [5, 6]. Currently, scRNA-seq is applied to develop personalized therapeutic strategies that are potentially useful in cancer diagnosis, therapy resistance during cancer progression, and the survival of patients [5, 7]. The scRNA-seq has also been adopted to combat COVID-19 to elucidate how the innate and adaptive host immune system miscommunicates resulting in worsening the immunopathology produced during this viral infection [8, 9].

These studies have led to a massive amount of scRNA-seq data deposited to public databases such as 10X Single-cell gene expression dataset, Human Cell Atlas, and Mouse Cell Atlas. Expressions of millions of cells from 18 species have been collected and deposited, waiting for further analysis. On the other hand, due to biological and technical factors, scRNA-seq data presents several analytical challenges related to its complex characteristics like missing expression values, high technical and biological variance, noise and sparse gene coverage, and elusive cell identities [1]. These characteristics make it difficult to directly apply commonly used bulk RNA-seq data analysis techniques and have called for novel statistical approaches for scRNA-seq data cleaning and computational algorithms for data analysis and interpretation. To this end, specialized scRNA-seq analysis pipelines such as Seurat [10] and Scanpy [11], along with a large collection of task-specific tools, have been developed to address the intricate technical and biological complexity of scRNA-seq data.

Recently, deep learning has demonstrated its significant advantages in natural language processing and speech and facial recognition with massive data [12-14]. Such advantages have initiated the application of DL in scRNA-seq data analysis as a competitive alternative to conventional machine learning approaches for uncovering cell clustering [15, 16], cell type identification [15, 17], gene imputation [18-20], and batch correction [21] in scRNA-seq analysis. Compared to conventional machine learning (ML) approaches, DL is more powerful in capturing complex features of high-dimensional scRNA-seq data. It is also more versatile, where a single model can be trained to address multiple tasks or adapted and transferred to different tasks. Moreover, the DL training scales more favorably with the number of cells in scRNA-seq data size, making it particularly attractive for handling the ever-increasing volume of single cell data. Indeed, the growing body of DL-based tools has demonstrated DL's exciting potential as a learning paradigm to significantly advance the tools we use to interrogate scRNA-seq data.

In this paper, we present a comprehensive review of the recent advances of DL methods for solving the present challenges in scRNA-seq data analysis (Table 1) from the quality control, normalization/batch effect reduction, dimension reduction, visualization, feature selection, and data interpretation by surveying deep learning papers published up to April 2021. In order to maintain high quality for this review, we choose not to include any (bio)archival papers, although a proportion of these manuscripts contain important new findings that would be published after completing their peer-reviewed process. Previous efforts to review the recent advances in machine learning methods focused on efficient integration of single cell data [22, 23]. A recent review of DL applications on single cell data has summarized 21 DL algorithms that might be deployed in single cell studies [24]. It also evaluated the clustering and data correction effect of these DL algorithms using 11 datasets.

In this review, we focus more on the DL algorithms with a much detailed explanation and comparison. Further, to better understand the relationship of each surveyed DL model with the overall scRNA-seq analysis pipeline, we organize the surveys according to the challenge they address and discuss these DL models following the analysis pipeline. A unified mathematical description of the surveyed DL models is presented and the specific model features are discussed when reviewing each method. This will also shed light on the modeling connections among the surveyed DL methods and the recognization of the uniqueness of each model. Besides the models, we also summarize the evaluation matrics used by these DL algorithms and methods that each DL algorithm was compared with. The online location of the code, the development platform, the used datasets for each method are also cataloged to facilitate their utilization and additional effort to improve them. Finally, we also created a companion online version of the paper at https://huang-ai4medicine-lab.github.io/survey-of-DL-for-scRNA-seq-analysis/gitbook/_book, which includes expanded discussion as well as a survey of additional methods. We envision that this survey will serve as an important information portal for learning the application of DL for scRNA-seq analysis and inspire innovative use of DL to address a broader range of new challenges in emerging multi-omics and spatial single-cell sequencing.

## 2. Overview of the scRNA-seq processing pipeline

Various scRNA-seq techniques (like SMART-seq, Drop-seq, and 10X genomics sequencing [25, 26] are available nowadays with their sets of advantages and disadvantages. Despite the differences in the scRNA-seq techniques, the data content and processing steps of scRNA-seq data are quite standard and conventional. A typical scRNA-seq dataset consists of three files: genes quantified (gene IDs), cells quantified (cellular barcode), and a count matrix (number of cells x number of genes), irrespective of the technology or pipeline used. A series of essential steps in scRNA-seq data processing pipeline and optional tools for each step with both ML and DL approaches are illustrated in **Fig. 1**.

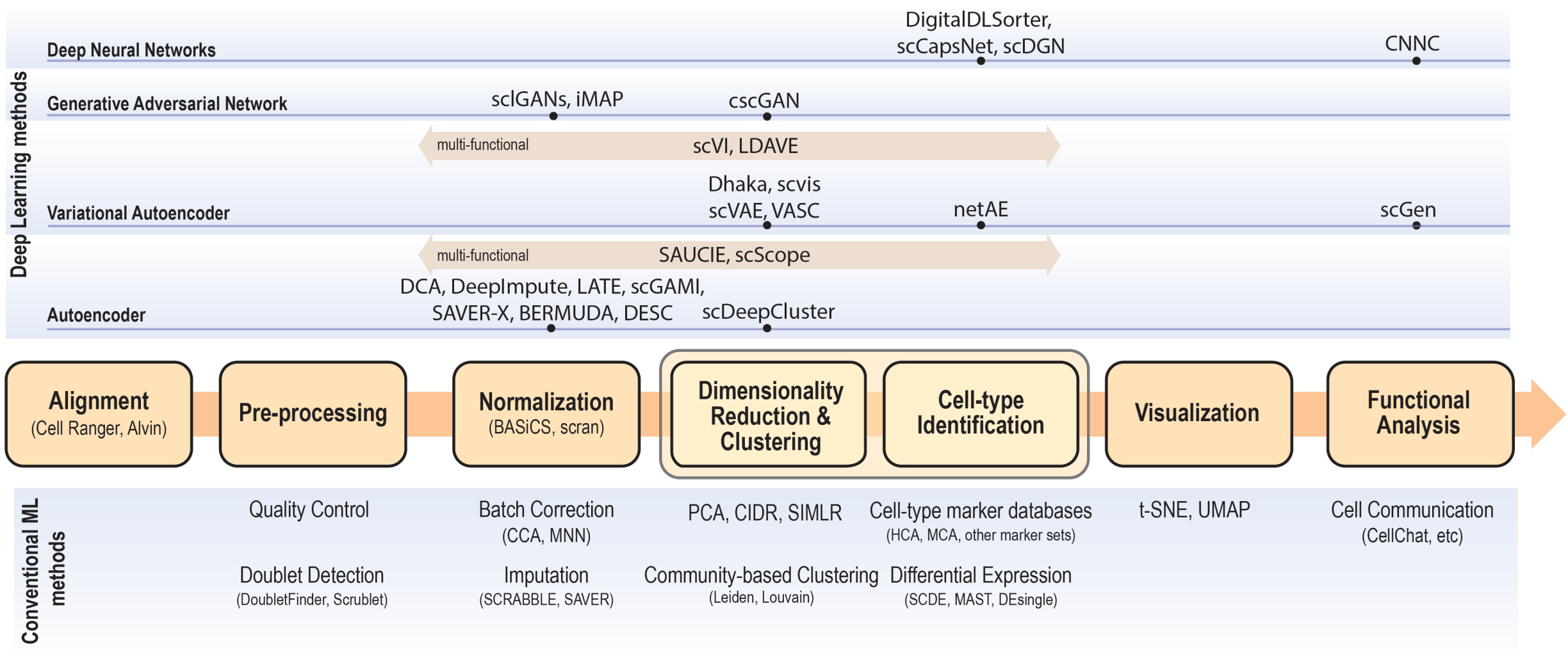

**Figure 1. Single cell data analysis steps for both conventional ML methods (bottom) and DL methods (top).** Depending on the input data and analysis objectives, major scRNA-seq analysis steps are illustrated in the the center flow chart (color boxes) with convential ML approaches along with optional analysis modules below each analysis step. Deep learning approaches are categorized as Deep Neural Network, Generalive Adversarial Network, Variational Autoencoder, and Autoencoder. For each DL approach, optional algorithms are listed on top of each step.

With the advantage of identifying each cell and unique molecular identifiers (UMIs) for expressions of each gene in a single cell, scRNA-seq data are embedded with increased technical noise and biases [27]. **Quality control (QC)** is the first and the key step to filter out dead cells, double-cells, or cells with failed chemistry or other technical artifacts. The most commonly adopted three QC covariates include the number of counts (count depth) per barcode identifying each cell, the number of genes per barcode, and the fraction of counts from mitochondrial genes per barcode [28].

**Normalization** is designed to eliminate imbalanced sampling, cell differentiation, viability, and many other factors. Approaches tailored for scRNA-seq have been developed including the Bayesian-based method coupled with spike-in, or BASiCS [29], deconvolution approach, scran [30], and sctransfrom in Seurat where regularized Negative Binomial Regression was proposed [31]. Two important steps, batch correction and imputation, will be carried out if required by the analysis.

- **Batch Correction** is a common source of technical variation in high-throughput sequencing experiments due to variant experimental conditions such as technicians and experimental time, imposing a major challenge in scRNA-seq data analysis. Batch effect correction algorithms include detection of mutual nearest neighbors (MNNs) [32], canonical correlation analysis (CCA) with Seurat [33], and Harmony algorithm through cell-type representation [34].

- **Imputation** step is necessary to handle high sparsity data matrix, due to missing value or dropout in scRNA-seq data analysis. Several tools have been developed to "impute" zero values in scRNA-seq data, such as SCRABBLE [35], SAVER [36], and scImpute [37].

**Dimensionality reduction and visualization** are essential steps to represent biological meaningful variations and high dimensionality with significantly reduced computational cost. Dimensionality reduction methods, such as PCA, are widely used in scRNA-seq data analysis to achieve that purpose. More advanced nonlinear approaches that preserve the topological structure and avoid overcrowding in lower dimension representation, such as LLE [38] (used in SLICER [39]), tSNE [40], and UMAP [41], have also been developed and adopted as a standard in single-cell data visualization.

**Clustering analysis** is a key step to identify cell subpopulations or distinct cell types to unravel the extent of heterogeneity and their associated cell-type-specific markers. Unsupervised clustering is frequently used here to categorize cells into clusters by their similarity often taken the aforementioned dimensionality-reduced representations as input, such as community detection algorithm Louvain [42] and Leiden [43], or data-driven dimensionality reduction followed with k-Means cluster by SIMLR [44].

**Feature selection** is another important step in single-cell RNA-seq analysis to select a subset of genes, or features, for cell-type identification and functional enrichment of each cluster. This step is achieved by differential expression analysis designed for scRNA-seq, such as MAST that used linear model fitting and likelihood ratio testing [45]; SCDE that adopted a Bayesian approach with a Negative Binomial model for gene expression and Poisson process for dropouts [46], or DEsingle that utilized a Zero-Inflated Negative Binomial model to estimate the dropouts [47].

Besides these key steps, downstream analysis can include cell type identification, coexpression analysis, prediction of perturbation response, where DL has also been applied. Other advanced analyses including trajectory inference and velocity and pseudotime analysis are not discussed here because most of the approaches on these topics are non-DL based.

## 3. Overview of common unsupervised deep learning models for scRNA-seq analysis

As batch correction, dimension reduction, imputation, and clustering are unsupervised learning tasks, we start our review by introducing the general formulations of variational autoencoder (VAE), the autoencoder (AE), or generative adversarial networks (GAN) for scRNA-seq together with their training strategies. We will focus on the different features of each method and bring attention to their uniqueness and novelty applications for scRNA-seq data in Section 4.

### 3.1. Variational Autoencoder

Let $\boldsymbol{x}_n$ represent a $G \times 1$ vector of gene expression (UMI counts or normalized, log-transformed expression) of $G$ genes in cell $n$, where $p\left(x_{gn} \middle| \nu_{gn}, \alpha_{gn}\right)$ follows some distribution (e.g., zero-inflated negative binomial (ZINB) or Gaussian), where $\nu_{gn}$ and $\alpha_{gn}$ are distribution parameters (e.g., mean, variance, or dispersion) (**Fig. 2A**). We consider $\nu_{gn}$ to be of particular interest (e.g., the mean counts) and is thus further modeled by a decoder neural network $D_{\boldsymbol{\theta}}$ (**Fig. 2A**) as

$$\boldsymbol{\nu}_n = D_{\boldsymbol{\theta}}(\boldsymbol{z}_n, s_n) \tag{1}$$

where the $g$th element of $\boldsymbol{\nu}_n$ is $\nu_{gn}$ and $\boldsymbol{\theta}$ is a vector of decoder weights, $\boldsymbol{z}_n \in \mathbb{R}^d$ represents a latent representation of gene expression and is used for visualization and clustering and $s_n$ is an observed variable (e.g., the batch ID). For VAE, $\boldsymbol{z}_n$ is commonly assumed to follow a multivariate standard normal prior, i.e., $p(\boldsymbol{z}_n) = \mathcal{N}(0, \mathbf{I}_d)$ with $\mathbf{I}_d$ being a $d \times d$ identity matrix. Further, $\alpha_{gn}$ of $p(x_{gn} | \nu_{gn}, \alpha_{gn})$ is a nuisance parameter, which has a prior distribution $p(\alpha_{gn})$ and can be either estimated or marginalized in variational inference. Now define $\boldsymbol{\Theta} = \{\boldsymbol{\theta}, \alpha_{ng}\ \forall n, g\}$. Then, $p(x_{gn} | \nu_{gn}, \alpha_{gn})$ and (1) together define the likelihood $p(\boldsymbol{x}_n | \boldsymbol{z}_n, s_n, \boldsymbol{\Theta})$.

The goal of training is to compute the maximum likelihood estimate of $\boldsymbol{\Theta}$

$$\widehat{\boldsymbol{\Theta}}_{ML} = \operatorname{argmax}_{\boldsymbol{\Theta}} \sum_{n=1}^{N} \log p(\boldsymbol{x}_n | s_n, \boldsymbol{\Theta}) \approx \operatorname{argmax}_{\boldsymbol{\Theta}} \sum_{n=1}^{N} \mathcal{L}(\boldsymbol{\Theta}), \tag{2}$$

where $\mathcal{L}(\boldsymbol{\Theta})$ is the evidence lower bound (ELBO),

$$\mathcal{L}(\boldsymbol{\Theta}) = \mathrm{E}_{q(\boldsymbol{z}_n | \boldsymbol{x}_n, s_n, \boldsymbol{\Theta})}[\log p(\boldsymbol{x}_n | \boldsymbol{z}_n, s_n, \boldsymbol{\Theta})] - D_{KL}[q(\boldsymbol{z}_n | \boldsymbol{x}_n, s_n, \boldsymbol{\Theta}) \| p(\boldsymbol{z}_n)], \tag{3}$$

and $q(\boldsymbol{z}_n|\boldsymbol{x}_n, s_n)$ is an approximate to $p(\boldsymbol{z}_n|\boldsymbol{x}_n, s_n)$ and assumed as

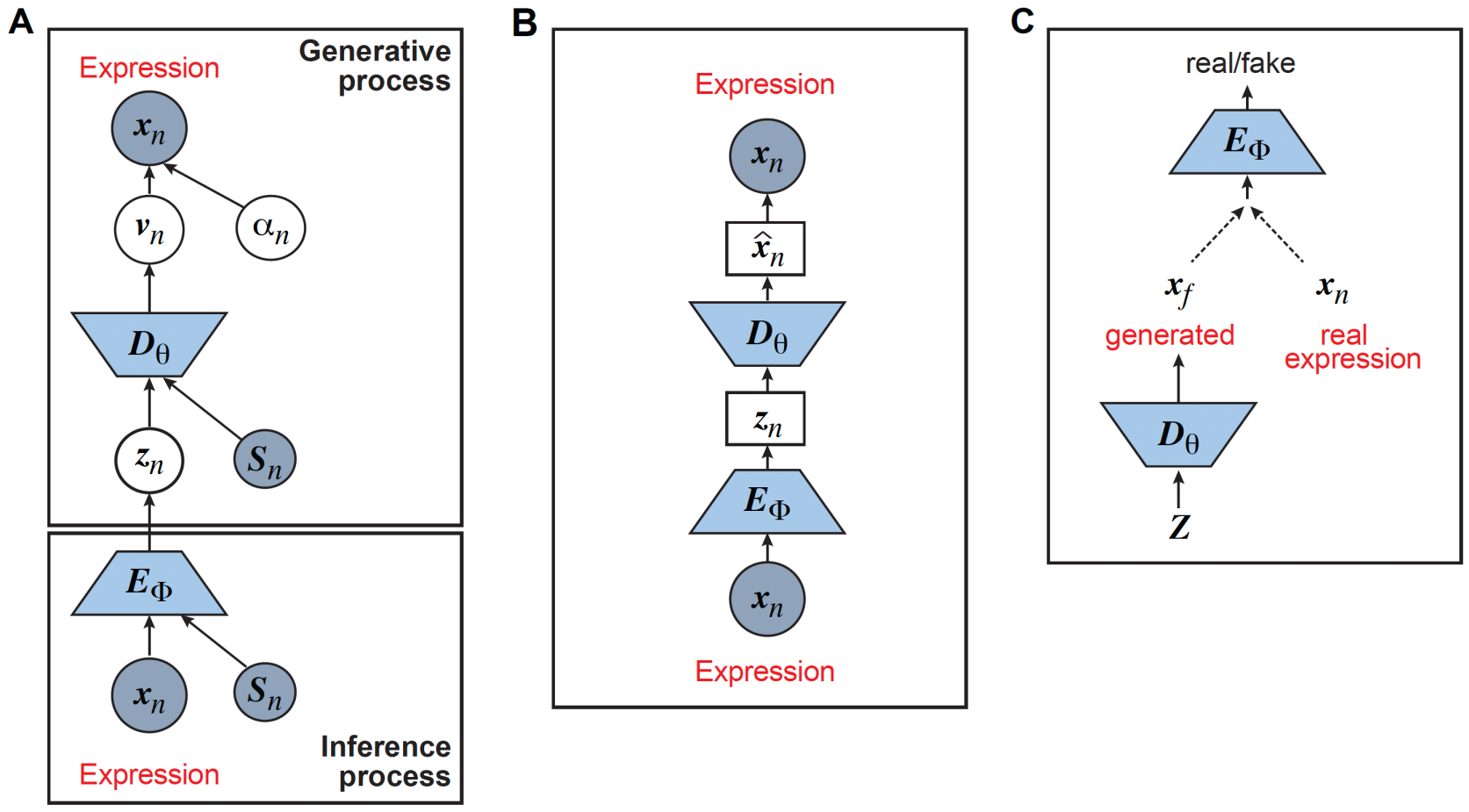

**Figure 2. Graphical models of the surveyed DL models including A)** Variational Autoencoder; **B)** Autoencoder; and **C)** Generative Adversarial Network.

$$q(\boldsymbol{z}_n|\boldsymbol{x}_n, s_n) = \mathcal{N}\left(\boldsymbol{\mu}_{z_n}, diag(\boldsymbol{\sigma}^2_{Z_n})\right), \tag{4}$$

with $\{\boldsymbol{\mu}_{z_n}, \boldsymbol{\sigma}^2_{Z_n}\}$ given by an encoder network $E_{\boldsymbol{\phi}}$ (**Fig. 2A**) as

$$\{\boldsymbol{\mu}_{z_n}, \boldsymbol{\sigma}^2_{Z_n}\} = E_{\boldsymbol{\phi}}(\boldsymbol{x}_n, s_n), \tag{5}$$

where $\boldsymbol{\phi}$ is the weights vector. Now, $\boldsymbol{\Theta} = \{\boldsymbol{\theta}, \boldsymbol{\phi}, \alpha_{ng}\ \forall n, g\}$ and equation (2) is solved by the stochastic gradient descent approach while a model is trained.

All the surveyed papers that deploy VAE follow this general modeling process. However, a more general formulation has a loss function defined as

$$L(\boldsymbol{\Theta}) = -\mathcal{L}(\boldsymbol{\Theta}) + \sum_{k=1}^{K} \lambda_k L_k(\boldsymbol{\Theta}), \tag{6}$$

where $L_k \forall k = 1, \ldots, K$ are losses for different functions (clustering, cell type prediction, etc) and $\lambda_k$s are the Lagrange multipliers. With this general formulation, for each paper, we examined the specific choices of data distribution $p(x_{gn}|\, v_{gn}, \alpha_{gn})$ that define $\mathcal{L}(\boldsymbol{\Theta})$, different $L_k$ designed for specific functions, and how the decoder and encoder were applied to model different aspects of scRNA-seq data.

### 3.2. Autoencoders

AEs learn the low dimensional latent representation $\boldsymbol{z}_n \in \mathbb{R}^d$ of expression $\boldsymbol{x}_n$. The AE includes an encoder $E_{\boldsymbol{\phi}}$ and a decoder $D_{\boldsymbol{\theta}}$ (**Fig. 2B**) such that

$$\boldsymbol{z}_n = E_{\boldsymbol{\phi}}(\boldsymbol{x}_n);\ \hat{\boldsymbol{x}}_n = D_{\boldsymbol{\theta}}(\boldsymbol{z}_n), \tag{7}$$

where $\boldsymbol{\Theta} = \{\boldsymbol{\theta}, \boldsymbol{\phi}\}$ are encoder and decoder weight parameters and $\hat{\boldsymbol{x}}_n$ defines the parameters (e.g. mean) of the likelihood $p(\boldsymbol{x}_n|\boldsymbol{\Theta})$ (**Fig. 2B**) and is often considered as imputed and denoised expressions. Additional design can be included in the AE model for batch correction, clustering, and other objectives.

The training of the AE is generally carried out by stochastic gradient descent algorithms to minimize the loss similar to Eq. (6) except $\mathcal{L}(\boldsymbol{\Theta}) = -\log p(\boldsymbol{x}_n|\boldsymbol{\Theta})$. When $p(\boldsymbol{x}_n|\boldsymbol{\Theta})$ is the Gaussian, $\mathcal{L}(\boldsymbol{\Theta})$ becomes the mean square error (MSE) loss

$$\mathcal{L}(\boldsymbol{\Theta}) = \sum_{n=1}^{N} \|\boldsymbol{x}_n - \hat{\boldsymbol{x}}_n\|_2^2. \tag{8}$$

Because different AE models differ in their AE architectures and the loss functions, we will discuss the specific architecture and the loss functions for each reviewed DL model in Section 4.

### 3.3. Generative adversarial networks

GANs have been used for imputation, data generation, and augmentation of the scRNA-seq analysis. Without loss of generality, the GAN, when applied to scRNA-seq, is designed to learn how to generate gene expression profiles from $p_x$, the distribution of $\boldsymbol{x}_n$. The vanilla GAN consists of two deep neural networks [48]. The first network is the generator $G_{\boldsymbol{\theta}}(\boldsymbol{z}_n, y_n)$ with parameter $\boldsymbol{\theta}$, a noise vector $\boldsymbol{z}_n$ from the distribution $p_z$ and a class label $y$ (e.g. cell type) and is trained to generate $\boldsymbol{x}_f$, a "fake" gene expression (**Fig.**

**2C**). The second network is the discriminator network $D_{\boldsymbol{\phi}_\mathrm{D}}$ with parameters $\boldsymbol{\phi}_\mathrm{D}$, trained to distinguish the "real" $\boldsymbol{x}$ from fake $\boldsymbol{x}_f$ (**Fig. 2C**). Both networks, $G_{\boldsymbol{\theta}}$ and $D_{\boldsymbol{\phi}_\mathrm{D}}$ are trained to outplay each other, resulting in a minimax game, in which $G_{\boldsymbol{\theta}}$ is forced by $D_{\boldsymbol{\phi}_\mathrm{D}}$ to produce better samples, which, when converge, can fool the discriminator $D_{\boldsymbol{\phi}_\mathrm{D}}$, thus becoming samples from $p_x$. The vanilla GAN suffers heavily from training instability and mode collapsing[49]. To that end, Wasserstein GAN (WGAN) [49] was developed with the WGAN loss [50]:

$$L(\boldsymbol{\theta}) = \max_{\phi_\mathrm{D}} \sum_{n=1}^{N} D_{\phi_\mathrm{D}}(\boldsymbol{x}_n) - \sum_{n=1}^{N} D_{\phi_\mathrm{D}}\left(G_\theta(\boldsymbol{z}_n, y_n)\right). \tag{9}$$

Additional terms can also be added to equation (9) to constrain the functions of the generator. Training based on the WGAN loss in Eq. (9) amounts to a min-max optimization, which iterates between the discriminator and generator, where each optimization is achieved by a stochastic gradient descent algorithm through backpropagation. The WGAN requires$D_{\boldsymbol{\phi}_\mathrm{D}}$ to be K-Lipschitz continuous [50], which can be satisfied by adding the gradient penalty to the WGAN loss [49]. Once the training is done, the generator $G_{\phi_G}$ can be used to generate gene expression profiles of new cells.

## 4. Survey of deep learning models for scRNA-seq analysis

In this section, we survey applications of DL models for scRNA-seq analysis. To better understand the relationship between the problems that each surveyed work addresses and the key challenges in the general scRNA-seq processing pipeline, we divide the survey into sections according to steps in the scRNA-seq processing pipeline illustrated in **Fig. 1**. For each DL model, we present the model details under the general model framework introduced in Section 3 and discuss the specific loss functions. We also survey the evaluation metrics and summarize the evaluation results. To facilitate cross-references of the information, we summarized all algorithms reviewed in this section in **Table 1** and tabulate the datasets and evaluation metrics used in each paper in **Tables 2 & 3.** We also listed all other algorithms that each surveyed method evaluated against in **Fig. 3**, highlighting the extensiveness these algorithms were assessed for their performance.

## 4.1. Imputation

### 4.1.1. DCA: deep count autoencoder

DCA [18] is an AE for imputation (**Figs. 2B, 4B**) and has been integrated into the Scanpy framework.

*Model*. DCA models UMI counts with missing values using the ZINB distribution

$$p(x_{gn} | \mathbf{\Theta}) = \pi_{gn}\delta(0) + (1 - \pi_{gn})NB(\nu_{gn}, \alpha_{gn}), \text{ for } g = 1, \dots G;\ n = 1, \dots N, \tag{10}$$

where $\delta(\cdot)$ is a Dirac delta function, $NB(\cdot,\cdot)$ denotes the negative binomial distribution, and $\pi_{gn}, \nu_{gn}, \alpha_{gn}$, representing dropout rate, mean, and dispersion, respectively, are functions of the output ($\hat{\boldsymbol{x}}_n$) of the decoder in the DCA as follows,

$$\boldsymbol{\pi}_n = sigmoid(\boldsymbol{W}_\pi \hat{\boldsymbol{x}}_n);\ \boldsymbol{\nu}_n = \exp(\boldsymbol{W}_\nu \hat{\boldsymbol{x}}_n);\ \boldsymbol{\alpha}_n = \exp(\boldsymbol{W}_\alpha \hat{\boldsymbol{x}}_n), \tag{11}$$

where $\boldsymbol{W}_\pi$, $\boldsymbol{W}_\nu$, and $\boldsymbol{W}_\alpha$ are additional weights to be estimated. The DCA encoder and decoder follow the general AE formulation as in Eq. (7) but the encoder takes the normalized, log-transformed expression as input. To train the model, DCA uses a constrained log-likelihood as the loss function

$$L(\mathbf{\Theta}) = \sum_{n=1}^{N}\sum_{g=1}^{G}\left(-log p\left(x_{gn} \mid \mathbf{\Theta}\right) + \lambda\pi_{gn}^{2}\right), \quad (12)$$

with $\mathbf{\Theta} = \{\boldsymbol{\theta}, \boldsymbol{\phi}, \boldsymbol{W}_{\pi}, \boldsymbol{W}_{v}, \boldsymbol{W}_{\alpha}\}$. Once the DCA is trained, the mean counts $\boldsymbol{v}_n$ are used as the denoised and imputed counts for cell $n$.

*__Results__*. For evaluation, DCA was compared to other methods using simulated data (using Splatter R package), and real bulk transcriptomics data from a developmental *C. elegans* time-course experiment was used with added simulating single-cell specific noise. Gene expression was measured from 206 developmentally synchronized young adults over a twelve-hour period (*C. elegans*). Single-cell specific noise was added *in silico* by genewise subtracting values drawn from the exponential distribution such that 80% of values were zeros. The paper analyzed the Bulk contains less noise than single-cell transcriptomics data and can thus aid in evaluating single-cell denoising methods by providing a good ground truth model. The authors also did a comparison of other methods including SAVER [36], scImpute [37], and MAGIC[51]. DCA denoising recovered original time-course gene expression pattern while removing single-cell specific noise. Overall, DCA demonstrated the strongest recovery of the top 500 genes most strongly associated with development in the original data without noise; DCA was shown to outperform other existing methods in capturing cell population structure in real data using PBMC, CITE-seq, runtime scales linearly with the number of cells.

### 4.1.2. SAVER-X: single-cell analysis via expression recovery harnessing external data

SAVER-X [52] is an AE model (**Figs. 2B, 4B**) developed to denoise and impute scRNA-seq data with transfer learning from other data resources.

*Model*. SAVER-X decomposes the variation in the observed counts $\boldsymbol{x}_n$ with missing values into three components: i) predictable structured component representing the shared variation across genes, ii) unpredictable cell-level biological variation and gene-specific dispersions, and iii) technical noise. Specifically, $x_{gn}$ is modeled as a Poisson-Gamma hierarchical model,

$$p(x_{gn}|\,\boldsymbol{\Theta}) = Poisson(l_n x'_{gn}), \qquad p(x'_{gn}|\nu_{gn}, \alpha_g) = Gamma(\nu_{gn}, \alpha_g \nu_{gn}^2), \tag{13}$$

where $l_n$ is the sequencing depth of cell $n$, $\nu_{gn}$ is the mean, and $\alpha_g$ is the dispersion. This Poisson-Gamma mixture is an equivalent expression to the NB distribution and thus, the ZINB distribution as Eq. (10) is adopted to model missing values.

The loss is similar to Eq. (12). However, $\nu_{gn}$ is initially learned by an AE pre-trained using external datasets from an identical or similar tissue and then transferred to $\boldsymbol{x}_n$ to be denoised. Such transfer learning can be applied to data between species (e.g., human and mouse in the study), cell types, batches, and single-cell profiling technologies. After $\nu_{gn}$ is inferred, SAVER-X generates the final denoised data $\hat{x}_{gn}$ by an empirical Bayesian shrinkage.

*Results.* SAVER-X was applied to multiple human single-cell datasets of different scenarios: i) T-cell subtypes, ii) a cell type (CD4+ regulatory T cells) that was absent from the pretraining dataset, iii) gene-protein correlations of CITE-seq data, and iv) immune cells of primary breast cancer samples with a pretraining on normal immune cells. SAVER-X with pretraining on HCA and/or PBMCs outperformed the same model without pretraining and other denoising methods, including DCA [28], scVI[17], scImpute [37], and MAGIC [51]. The model achieved promising results even for genes with very low UMI counts. SAVER-X was also applied for a cross-species study in which the model was pre-trained on a human or mouse dataset and transferred to denoise another. The results demonstrated the merit of transferring public data resources to denoise in-house scRNA-seq data even when the study species, cell types, or single-cell profiling technologies are different.

### 4.1.3. DeepImpute: Deep neural network Imputation

DeepImpute [20] imputes genes in a divide-and-conquer approach, using a bank of DNN models (**Fig. 4A**) with 512 output, each to predict gene expression levels of a cell.

*Model.* For each dataset, DeepImpute selects to impute a list of genes or highly variable genes (variance over mean ratio, default = 0.5). Each sub-neural network aims to understand the relationship between the input genes and a subset of target genes. Genes are first divided into $N$ random subsets of 512 target genes. For each subset, a two-layer DNN is trained where the input includes genes that are among the top 5 best-correlated genes to target genes but not part of the target genes in the subset. The loss is defined as the weighted MSE

$$\mathcal{L}(\boldsymbol{\Theta}) = \sum \boldsymbol{x}_n(\boldsymbol{x}_n - \widehat{\boldsymbol{x}}_n)^2, \quad (14)$$

which gives higher weights to genes with higher expression values.

*Result.* DeepImpute had the highest overall accuracy and offered shorter computation time with less demand on computer memory than other methods like MAGIC, DrImpute, ScImpute, SAVER, VIPER, and DCA. Using simulated and experimental datasets (**Table 2**), DeepImpute showed benefits in improving clustering results and identifying significantly differentially expressed genes. DeepImpute and DCA, show overall advantages over other methods and between which DeepImpute performs even better. The properties of DeepImpute contribute to its superior performance include 1) a divide-and-conquer approach which contrary to an autoencoder as implemented in DCA, resulting in a lower complexity in each sub-model and stabilizing neural networks, and 2) the subnetworks are trained without using the target genes as the input which reduces overfitting while enforcing the network to understand true relationships between genes.

#### 4.1.4. LATE: Learning with AuToEncoder

LATE [53] is an AE (**Figs. 2B, 4B**) whose encoder takes the log-transformed expression as input.

*Model*. LATE sets zeros for all missing values and generates the imputed expressions. LATE minimizes the MSE loss as Eq. (8). One issue is that some zeros could be real and reflect the actual lack of expressions.

*Result.* Using synthetic data generated from pre-imputed data followed by random dropout selection at different degrees, LATE outperforms other existing methods like MAGIC, SAVER, DCA, scVI, particularly when the ground truth contains only a few or no zeros. However, when the data contain many zero expression values, DCA achieved a lower MSE than LATE, although LATE still has a smaller MSE than scVI. This result suggests that DCA likely does a better job identifying true zero expressions, partly because LATE does not make assumptions on the statistical distributions of the single-cell data that potentially have inflated zero counts.

#### 4.1.5. scGMAI

Technically, scGMAI [54] is a model for clustering but it includes an AE (**Figs. 2B, 4B**) in the first step to combat dropout.

*Model.* To impute the missing values, scGMAI applies an AE like LATE to reconstruct log-transformed expressions with dropout but chooses a smoother Softplus activation function instead. The MSE loss as in Eq. (8) is adopted.

After imputation, scGMAI uses fast independent component analysis (ICA) on the AE reconstructed expressions to reduce the dimension and then applies a Gaussian mixture model on the ICA reduced data to perform the clustering.

*Results.* To assess the performance, the AE in scGMAI was replaced by five other imputation methods including SAVER [36], MAGIC [51], DCA [28], scImpute [37], and CIDR[55]. A scGMAI implementation without AE was also compared. Seventeen scRNA-seq data (part of them are listed in **Tables 2b & c** as marked) were used to evaluate cell clustering performances. The results indicated that the AEs significantly improved the clustering performance in eight of seventeen scRNA-seq datasets.

#### 4.1.6. scIGANs

Imputation approaches based on information from cells with similar expressions suffer from oversmoothing, especially for rare cell types. scIGANs [19] is a GAN-based imputation algorithm (**Figs. 2C, 4E**), which overcomes this problem by training a GAN model to generate samples with imputed expressions.

*Model.* scIGAN takes the image-like reshaped gene expression data $\boldsymbol{x}_n$ as input. The model follows a BEGAN [56] framework, which replaces the GAN discriminator $D$ with a function $R_{\emptyset_R}$ to compute the reconstruction MSE. Then, the Wasserstein distance loss between the reconstruction errors between the real and generated samples are computed

$$L(\boldsymbol{\theta}, \boldsymbol{\Phi}) = \max_{\emptyset_R} \sum_{n=1}^{N} R_{\emptyset_R}(\boldsymbol{x}_n) - \sum_{n=1}^{N} R_{\emptyset_R}(G_\theta(E_\Phi(\boldsymbol{x}_n), y), \quad (15)$$

This framework forces the model to meet two computing objectives, i.e. reconstructing the real samples and discriminating between real and generated samples. Proportional Control Theory was applied to balance these two goals during the training [57].

After training, the decoder $G_\theta$ is used to generate new samples of a specific cell type. Then, the k-nearest neighbors (KNN) approach is applied to the real and generated samples to impute the real samples' missing expressions.

*Results.* scIGANs was first tested on simulated samples with different dropout rates. Performance of rescuing the correct clusters was compared with 11 existing imputation approaches including DCA, DeepImpute, SAVER, scImpute, MAGIC, etc. scIGANs reported the best performance for all metrics. scIGAN was next evaluated for its ability to correctly cluster cell types on the Human brain scRNA-seq data, which showed superior performance than existing methods again. scIGANs was then evaluated for identifying cell-cycle states using scRNA-seq datasets from mouse embryonic stem cells. The results showed that scIGANs outperformed competing existing approaches for recovering subcellular states of cell cycle dynamics. scIGANs were further shown to improve the identification of differentially expressed genes and enhance the inference of cellular trajectory using time-course scRNA-seq data from the differentiation from H1 ESC to definitive endoderm cells (DEC). Finally, scIGAN was also shown to scale to scRNA-seq methods and data sizes.

## 4.2. Batch effect correction

### 4.2.1. BERMUDA: Batch Effect ReMoval Using Deep Autoencoders

BERMUDA [58] deploys a transfer-learning method (**Figs. 2B, 4B**) to remove the batch effect. It performs correction to the shared cell clusters among batches and therefore preserves batch-specific cell populations.

*Model.* BERMUDA is an AE that takes normalized, log-transformed expression as input. Its consists of two parts as

$$L(\mathbf{\Theta}) = \mathcal{L}(\mathbf{\Theta}) + \lambda L_{MMD}(\mathbf{\Theta}), \tag{16}$$

where $\mathcal{L}(\mathbf{\Theta})$ is the MSE loss and $L_{MMD}$ is the maximum mean discrepancy (MMD) [59] loss that measures the differences in distributions between pairs of similar cell clusters shared among batches as:

$$L_{MMD}(\mathbf{\Theta}) = \sum_{i_a, i_b, j_a, j_b} M_{i_a, j_a, i_b, j_b} MMD(\mathbf{z}_{i_a, j_a}, \mathbf{z}_{i_b, j_b}), \tag{17}$$

where $\mathbf{z}_{i,j}$ is the latent variable of $\mathbf{x}_{i,j}$, the input expression of a cell from cluster $j$ of batch $i$, $M_{i_a, j_a, i_b, j_b}$ is $1$ if cluster $i_a$ of batch $j_a$ and cluster $i_b$ of batch $j_b$ are determined to be similar by MetaNeighbor [60] and $0$, otherwise. The $MMD$ equals zero when the underlying distributions of the observed samples are the same.

*Results.* BERMUDA was shown to outperform other methods like mnnCorrect [32], BBKNN[61], Seurat [10], and scVI [17] in removing batch effects on simulated and human pancreas data while preserving batch-specific biological signals. BERMUDA provides several improvements compared to existing methods: 1) capable of removing batch effects even when the cell population compositions across different batches are vastly different; and 2) preserving batch-specific biological signals through transfer-learning which enables discovering new information that might be hard to extract by analyzing each batch individually.

### 4.2.2. DESC: batch correction based on clustering

DESC [62] is an AE model (**Figs. 2B, 4B**) that removes batch effect through clustering with the hypothesis that batch differences in expressions are smaller than true biological variations between cell types, and, therefore, properly performing clustering on cells across multiple batches can remove batch effects without the need to define batches explicitly.

*Model.* DESC has a conventional AE architecture. Its encoder takes normalized, log-transformed expression and uses decoder output, $\hat{\boldsymbol{x}}_n$ as the reconstructed gene expression, which is equivalent to a Gaussian data distribution with $\hat{\boldsymbol{x}}_n$ being the mean. The loss function is similar to Eq. (16) and except that the second loss $L_c$ is the clustering loss that regularizes the learned feature representations to form clusters as in the deep embedded clustering [63]. The model is first trained to minimize $\mathcal{L}(\boldsymbol{\Theta})$ only to obtain the initial weights before minimizing the combined loss. After the training, each cell is assigned with a cluster ID.

*Results.* DESC was applied to the macaque retina dataset, which includes animal level, region level, and sample-level batch effects. The results showed that DESC is effective in removing the batch effect, whereas CCA [33], MNN [32], Seurat 3.0 [10], scVI [17], BERMUDA [58], and scanorama [64] were all sensitive to batch definitions. DESC was then applied to human pancreas datasets to test its ability to remove batch effects from multiple scRNA-seq platforms and yielded the highest ARI among the comparing approaches mentioned above. When applied to human PBMC data with interferon-beta stimulation, where biological variations are compounded by batch effect, DESC was shown to be the best in removing batch effect while preserving biological variations.

DESC was also shown to remove batch effect for the monocytes and mouse bone marrow data and DESC was shown to preserve the pseudotemporal structure. Finally, DESC scales linearly with the number of cells, and its running time is not affected by the increasing number of batches.

#### 4.2.3. iMAP: Integration of Multiple single-cell datasets by Adversarial Paired-style transfer networks

iMAP [65] combines AE (**Figs. 2B, 4B**) and GAN (**Figs. 2C, 4E**) for batch effect removal. It is designed to remove batch biases while preserving dataset-specific biological variations.

<u>Model.</u> iMAP consists of two processing stages, each including a separate DL model. In the first stage, a special AE, whose decoder combines the output of two separate decoders $D_{\boldsymbol{\theta}_1}$ and $D_{\boldsymbol{\theta}_2}$, is trained such that

$$\boldsymbol{z}_n = E_{\boldsymbol{\phi}}(\boldsymbol{x}_n); \ \widehat{\boldsymbol{x}}_n = D_{\boldsymbol{\theta}}(\boldsymbol{z}_n, s_n) = ReLu(D_{\boldsymbol{\theta}_1}(s_n) + D_{\boldsymbol{\theta}_2}(\boldsymbol{z}_n, s_n)), \tag{18}$$

where $s_n$ is the one-hot encoded batch number of cell $n$. $D_{\boldsymbol{\theta}_1}$ can be understood as decoding the batch noise, whereas $D_{\boldsymbol{\theta}_2}$ reconstructs batch-removed expression from the latent variable $\boldsymbol{z}_n$. The training minimizes the loss in Eq. (16) except the 2nd loss is the content loss

$$L_t(\boldsymbol{\Theta}) = \sum_{n=1}^{N} \left\| \boldsymbol{z}_n - E_{\boldsymbol{\phi}}\left(D_{\boldsymbol{\theta}}(\boldsymbol{z}_n, \tilde{s}_n)\right) \right\|_2^2, \tag{19}$$

where $\tilde{s}_n$ is a random batch number. Minimizing $L_t(\boldsymbol{\Theta})$ further ensures the reconstructed expression $\widehat{\boldsymbol{x}}_n$ would be batch agnostic and has the same content as $\boldsymbol{x}_n$.

However, due to the limitation of AE, this step is still insufficient for batch removal. Therefore, a second stage is included to apply a GAN model to make expression distributions of the shared cell type across different baches indistinguishable. To identified the shared cell types, a mutual nearest neighbors (MNN) strategy adapted from [32] was developed to identify MNN pairs across batches using batch effect independent $\boldsymbol{z}_n$ as opposed to $\boldsymbol{x}_n$. Then, a mapping generator $G_{\boldsymbol{\theta}_G}$ is trained using MNN pairs based on GAN such that $\boldsymbol{x}_n^{(A)} = G_{\boldsymbol{\theta}_G}\left(\boldsymbol{x}_n^{(S)}\right)$, where $\boldsymbol{x}_n^{(S)}$ and $\boldsymbol{x}_n^{(A)}$ are the MNN pairs from batch $S$ and an anchor batch $A$. The WGAN-GP loss as in Eq. (9) was adopted for the GAN training. After training, $G_{\boldsymbol{\theta}_G}$ is applied to all cells of a batch to generate batch-corrected expression.

*Results*: iMAP was first tested on benchmark datasets from human dendritic cells and Jurkat and 293T cell lines and then human pancreas datasets from five different platforms. All the datasets contain both batch-specific cells and batch-shared cell types. iMAP was shown to separate the batch-specific cell types but mix batch shared cell types and outperformed 9 other existing batch correction methods including Harmony, scVI, fastMNN, Seurat, etc. iMAP was then applied to the large-scale Tabula Muris datasets containing over 100K cells sequenced from two platforms. iMAP could not only reliably integrate cells from the same tissues but identify cells from platform-specific tissues. Finally, iMAP was applied to datasets of tumor-infiltrating immune cells and shown to reduce the dropout ratio and the percentage of ribosomal genes and non-coding RNAs, thus improving detection of rare cell types and ligand-receptor interactions. iMAP scales with the number of cells, showing minimal time cost increase after the number of cells exceeds thousands. Its performance is also robust against model hyperparameters.

## 4.3. Dimension reduction, latent representation, clustering, and data augmentation

### 4.3.1. Dimension reduction by AEs with gene-interaction constrained architecture

This study [66] considers AEs (**Figs. 2B, 4B**) for learning the low-dimensional representation and specifically explores the benefit of incorporating prior biological knowledge of gene-gene interactions to regularize the AE network architecture.

*Model.* Several AE models with single or two hidden layers that incorporate gene interactions reflecting transcription factor (TF) regulations and protein-protein interactions (PPIs) are implemented. The models take normalized, log-transformed expressions and follow the general AE structure, including dimension-reducing and reconstructing layers, but the network architectures are not symmetrical. Specifically, gene interactions are incorporated such that each node of the first hidden layer represented a TF or a protein in the PPI; only genes that are targeted by TFs or involved in the PPI were connected to the node. Thus, the corresponding weights of $E_{\phi}$ and $D_{\theta}$ are set to be trainable and otherwise fixed at zero throughout the training process. Both unsupervised (AE-like) and supervised (cell-type label) learning were studied.

*Results.* Regularizing encoder connections with TF and PPI information considerably reduced the model complexity by almost 90% (7.5-7.6M to 1.0-1.1M). The clusters formed on the data representations learned from the models with or without TF and PPI information were compared to those from PCA, NMF, independent component analysis (ICA), t-SNE, and SIMLR [44]. The model with TF/PPI information and 2 hidden layers achieved the best performance by five of the six measures and the best average performance. In terms of the cell-type retrieval of single cells, the encoder models with and without TF/PPI information achieved the best performance in 4 and 3 cell types, respectively. PCA yielded the best performance in only 2 cell types. The DNN model with TF/PPI information and 2 hidden layers again achieved the best average performance across all cell types. In summary, this study demonstrated a biologically meaningful way to regularize AEs by the prior biological knowledge for learning the representation of scRNA-seq data for cell clustering and retrieval.

#### 4.3.2. Dhaka: a VAE-based dimension reduction model.

Dhaka [67] was proposed to reduce the dimension of scRNA-seq data for efficient stratification of tumor subpopulations.

_Model._ Dhaka adopts a general VAE formulation (**Figs. 2A, 4C**). It takes the normalized, log-transformed expressions of a cell as input and outputs the low-dimensional representation.

_Result._ Dhaka was first tested on the simulated dataset. The simulated dataset contains 500 cells, each including 3K genes, clustered into 5 different clusters with 100 cells each. The clustering performance was compared with other methods including t-SNE, PCA, SIMLR, NMF, an autoencoder, MAGIC, and scVI. Dhaka was shown to have an ARI higher than most other comparing methods. Dhaka was then applied to the Oligodendroglioma data and could separate malignant cells from non-malignant microglia/macrophage cells. It also uncovered the shared glial lineage and differentially expressed genes for the lineages. Dhaka was also applied to the Glioblastoma data and revealed an evolutionary trajectory of the malignant cells where cells gradually evolve from a stemlike state to a more differentiated state. In contrast, other methods failed to capture this underlying structure. Dhaka was next applied to the Melanoma cancer dataset [68] and uncovered two distinct clusters that showed the intra-tumor heterogeneity of the Melanoma samples. Dhaka was finally applied to copy number variation data [69] and shown to identify one major and one minor cell clusters, of which other methods could not find.

#### 4.3.3. scvis: a VAE for capturing low-dimensional structures

scvis [70] is a VAE network (**Figs. 2A, 4C**) that learns the low-dimensional representations capture both local and global neighboring structures in scRNA-seq data.

<u>*Model:*</u> scvis adopts the generic VAE formulation described in section 3.1. However, it has a unique loss function defined as

$$L(\boldsymbol{\Theta}) = -\mathcal{L}(\boldsymbol{\Theta}) + \lambda L_t(\boldsymbol{\Theta}), \tag{20}$$

where $\mathcal{L}(\boldsymbol{\Theta})$ is ELBO as in Eq. (3) and $L_t$ is a regularizer using non-symmetrized t-SNE objective function [70], which is defined as

$$L_t(\boldsymbol{\Theta}) = \sum_{i=1}^{N} \sum_{j=1, j\neq i}^{N} p_{j|i} \log \frac{p_{j|i}}{q_{j|i}}, \tag{21}$$

where $i$ and $j$ are two different cells, $p_{i|j}$ measures the local cell relationship in the data space, and $q_{j|i}$ measures such relationship in the latent space. Because t-SNE algorithm preserves the local structure of high dimensional space, $L_t$ learns local structures of cells.

<u>*Results.*</u> scvis was tested on the simulated data and outperformed t-SNE in a nine-dimensional space task. scvis preserved both local structure and global structure. The relative positions of all clusters were well kept but outliers were scattered around clusters. Using simulated data and comparing to t-SNE, scvis generally produced consistent and better patterns among different runs while t-SNE could not. scvis also presented good results on adding new data to an existing embedding, with median accuracy on new data at 98.1% for K= 5 and 94.8% for *K*= 65, when train *K* cluster on original data then test the classifier on new generated sample points. The scvis was subsequently tested on four real datasets including metastatic melanoma, oligodendroglioma, mouse bipolar and mouse retina datasets. In each dataset, scvis was showed to preserve both the global and local structure of the data.

#### 4.3.4. scVAE: VAE for single-cell gene expression data

scVAE [71] includes multiple VAE models (**Figs. 2A, 4C**) for denoising gene expression levels and learning the low-dimensional latent representation of cells. It investigates different choices of the likelihood functions in the VAE model to model different data sets.

_Model._ scVAE is a conventional fully connected network. However, different distributions have been discussed for $p(x_{gn} | \nu_{gn}, \alpha_{gn})$ to model different data behaviors. Specifically, scVAE considers Poisson, constrained Poisson, and negative binomial distributions for count data, piece-wise categorical Poisson for data including both high and low counts, and zero-inflated version of these distributions to model missing values. To model multiple modes in cell expressions, a Gaussian mixture is also considered for $q(\boldsymbol{z}_n | \boldsymbol{x}_n, s_n)$, resulting a GMVAE. The inference process still follows that of a VAE as discussed in section 3.1.

_Results._ scVAEs were evaluated on the PBMC data and compared with factor analysis (FA) models. The results showed that GMVAE with negative binomial distribution achieved the highest lower bound and ARI. Zero-inflated Poisson distribution performed the second best. All scVAE models outperformed the baseline linear factor analysis model, which suggested that a non-linear model is needed to capture single-cell genomic features. GMVAE was also compared with Seurat and shown to perform better using the withheld data. However, scVAE performed no better than scVI [17] or scvis [70], both are VAE models.

#### 4.3.5. VASC: VAE for scRNA-seq

VASC [72] is another VAE (**Figs. 2A, 4C**) for dimension reduction and latent representation but it models dropout.

*Model:* VASC's input is the log-transformed expression but rescaled in the range [0,1]. A dropout layer (dropout rate of 0.5) is added after the input layer to force subsequent layers to learn to avoid dropout noise. The encoder network has three layers fully connected and the first layer uses linear activation, which acts like an embedded PCA transformation. The next two layers use the ReLU activation, which ensures a sparse and stable output. This model's novelty is the zero-inflation layer (ZI layer), which is added after the decoder to model scRNA-seq dropout events. The probability of dropout event is defined as $e^{-\hat{x}^2}$ where $\hat{x}$ is the recovered expression value obtained by the decoder network. Since backpropagation cannot deal with a stochastic network with categorical variables, a Gumbel-softmax distribution [73] is introduced to address the difficulty of the ZI layer. The loss function of the model takes the form $L = \mathcal{L}(\mathbf{\Theta}) + \lambda L_{KL}(\mathbf{\Theta})$, where $\mathcal{L}$ is the binary entropy because the input is scaled to [0 1], and$L_{KL}$ a loss performed using KL divergence on the latent variables. After the model is trained, the latent code can be used as the dimension-reduced feature for downstream tasks and visualization.

*Results.* VASC was compared with PCA, t-SNE, ZIFA, and SIMLR on 20 datasets. In the study of embryonic development from zygote to blast cells, all methods roughly re-established the development stages of different cell types in the dimension-reduced space. However, VASC showed better performance to model embryo developmental progression. In the Goolam, Biase and Yan datasets, scRNA-seq data were generated through embryonic development stages from zygote to blast, VASC re-established development stage from 1, 2, 4, 8, 16 to blast, while other methods failed. In the Pollen, Kolodziejczyk, and Baron dataset, VASC formed an appropriate cluster, either with homogeneous cell type, preserved proper relative positions, or minimal batch influence. Interestingly, when tested on the PBMC dataset, VASC was showed to identify the major global structure (B cells, CD4+, CD8+ T cells, NK cells, Dendritic cells), and detect subtle differences within monocytes (FCGR3A+ vs CD14+ monocytes), indicating the capability of VASC handling a large number of cells or cell types. Quantitative clustering performance in NMI, ARI, homogeneity and completeness was also performed. VASC always ranked top two in all the datasets. In terms of NMI and ARI, VASC best performed on 15 and 17 out of 20 datasets, respectively.

### 4.3.6. scDeepCluster

scDeepCluster [74] is an AE network (**Figs. 2B, 4B**) that simultaneously learns feature representation and performs clustering via explicit modeling of cell clusters as in DESC.

_Model:_ Similar to DCA, scDeepCluster adopts a ZINB distribution for $\boldsymbol{x}_n$ as in Eq. (10) and (12). The loss is similar to Eq. (16) but with the first being the negative log-likelihood of the ZINB data distribution as defined in Eq. (12) and the second $L_c$ being a clustering loss performed using KL divergence as in DESC algorithm. Compared to csvis, scDeepcluster focuses more on clustering assignment due to the KL divergence.

_Results._ scDeepCluster was first tested on the simulation data and compared with other seven methods including DCA [18], two multi-kernel spectral clustering methods MPSSC [75] and SIMLR [44], CIDR [55], PCA + k-mean, scvis [70] and DEC[76]. In different dropout rate simulations, scDeepCluster significantly outperformed the other methods consistently. In signal strength, imbalanced sample size, and scalability simulations, scDeepcluster outperformed all other algorithms and scDeepCluster and most notably advantages for weak signals, robust against different data imbalance levels and scaled linearly with the number of cells. scDeepCluster was then tested on four real datasets (10X PBMC, Mouse ES cells, Mouse bladder cells, Worm neuron cells) and shown to outperform all other comparing algorithms. Particularly, MPSSC and SIMLR failed to process the full datasets due to quadratic complexity.

#### 4.3.7. cscGAN: Conditional single-cell generative adversarial neural networks

cscGAN [77] is a GAN model (**Figs. 2C, 4E**) designed to augment the existing scRNA-seq samples by generating expression profiles of specific cell types or subpopulations.

*Model.* Two models, csGAN and cscGAN, were developed following the general formulation of WGAN described in section 3.3. The difference between the two models is that cscGAN is a conditional GAN such that the input to the generator also includes a class label $y$ or cell type, i.e. $\boldsymbol{\phi}_G(\boldsymbol{z}, y)$. The projection-based conditioning (PCGAN) method [78] was adopted to obtain the conditional GAN. For both models, the generator (three layers of 1024, 512, and 256 neurons) and discriminator (three layers of 256, 512, and 1024 neurons) are fully connected DNNs.

*Results:* The performance of scGAN was first evaluated using PBMC data. The generated samples were shown to capture the desired clusters and the real data's regulons. Additionally, the AUC performance for classifying real from generated samples by a Random Forest classifier only reached 0.65, performance close to 0.5. Finally, scGAN's generated samples had a smaller MMD than those of Splatter, a state-of-the-art scRNA-seq data simulator [79]. Even though a large MMD was observed for scGAN when compared with that of SUGAR, another scRNA-seq simulator, SUGAR [80] was noted for prohibitively high runtime and memory. scGAN was further trained and assessed on the bigger mouse brain data and shown to model the expression dynamics across tissues. Then, the performance of cscGAN for generating cell-type-specific samples was evaluated using the PBMC data. cscGAN was shown to generate high-quality scRAN-seq data for specific cell types. Finally, the real PBMC samples were augmented with the generated samples. This augmentation improved the identification of rare cell types and the ability to capture transitional cell states from trajectory analysis.

### 4.4. Multi-functional models

Given the versatility of AE and VAE in addressing different scRAN-seq analysis challenges, DL models possessing multiple analysis functions have been developed. We survey these models in this section.

#### 4.4.1. scVI: single-cell variational inference

scVI [17] is designed to address a range of fundamental analysis tasks, including batch correction, visualization, clustering, and differential expression.

*Model.* scVI is a VAE (**Figs. 2A, 4C**) that models the counts of each cell from different batches. scVI adopts a ZINB distribution for $x_{gn}$

$$p\left(x_{gn} \middle| \pi_{gn}, L_n, \nu_{gn}, \alpha\right) = \pi_{gn}\delta(0) + \left(1 - \pi_{gn}\right)NB\left(L_n \nu_{gn}, \alpha_g\right), \tag{22}$$

which is defined similarly as Eq (11) in DCA, except that $L_n$ denotes the scaling factor for cell $n$, which follows a log-Normal ($log\mathcal{N}$) prior as $p(L_n) = log\mathcal{N}\left(\mu_{L_n}, \sigma_{L_n}^2\right)$, therefore, $\nu_{gn}$ represents the mean counts normalized by $L_n$. Now, let $s_n \in \{0,1\}^B$ be the batch ID of cell $n$ with $B$ being the total number of batches. Then, $\nu_{gn}$ and $\pi_g$ are further modeled as functions of the $d$-dimension latent variable $\boldsymbol{z}_n \in \mathbb{R}^d$ and the batch ID $s_n$ by the decoder networks $D_{\boldsymbol{\theta}_\nu}$ and $D_{\boldsymbol{\theta}_\pi}$ as

$$\boldsymbol{\nu}_n = D_{\boldsymbol{\theta}_\nu}(\boldsymbol{z}_n, s_n),\ \ \boldsymbol{\pi}_n = D_{\boldsymbol{\theta}_\pi}(\boldsymbol{z}_n, s_n), \tag{23}$$

where the $g$th element of $\boldsymbol{\nu}_n$ and $\boldsymbol{\pi}_n$ are $\nu_{gn}$ and $\pi_g$, respectively, and $\boldsymbol{\theta}_\nu$, and $\boldsymbol{\theta}_\pi$ are the decoder weights. Note that the lower layers of the two decoders are shared. For inference, both $\boldsymbol{z}_n$ and $L_n$ are considered as latent variables and therefore $q(x_n, s_n) = q(\boldsymbol{z}_n|\boldsymbol{x}_n, s_n)q(L_n|\boldsymbol{x}_n, s_n)$ is a mean-field approximate to the intractable posterior distribution $p(\boldsymbol{z}_n, L_n|\boldsymbol{x}_n, s_n)$ and

$$\begin{aligned} q(\boldsymbol{z}_n|\boldsymbol{x}_n, s_n) &= \mathcal{N}\left(\boldsymbol{\mu}_{z_n}, diag(\boldsymbol{\sigma}^2_{Z_n})\right), \\ q(L_n|\boldsymbol{x}_n, s_n) &= log\mathcal{N}\left(\boldsymbol{\mu}_{L_n}, diag(\boldsymbol{\sigma}^2_{L_n})\right), \end{aligned} \tag{24}$$

whose means and variances $\{\boldsymbol{\mu}_{z_n}, \boldsymbol{\sigma}^2_{Z_n}\}$ and $\{\boldsymbol{\mu}_{L_n}, \boldsymbol{\sigma}^2_{L_n}\}$ are defined by the encoder networks $E_Z$ and $E_L$ applied to $\boldsymbol{x}_n$ and $s_n$ as

$$\begin{aligned} \{\boldsymbol{\mu}_{z_n}, \boldsymbol{\sigma}^2_{Z_n}\} &= E_{\boldsymbol{\phi}_z}(\boldsymbol{x}_n, s_n), \\ \{\boldsymbol{\mu}_{L_n}, \boldsymbol{\sigma}^2_{L_n}\} &= E_{\boldsymbol{\phi}_L}(\boldsymbol{z}_n, s_n) \end{aligned} \tag{25}$$

where $\boldsymbol{\phi}_z$, and $\boldsymbol{\phi}_L$ are the encoder weights. Note that, like the decoders, the lower layers of the two encoders are also shared. Overall, the model parameters to be estimated by the variational optimization is $\boldsymbol{\Theta} = \{\boldsymbol{\theta}_\nu, \boldsymbol{\theta}_\pi, \boldsymbol{\phi}_z, \boldsymbol{\phi}_L, \alpha_g\}$. After inference, $\boldsymbol{z}_n$ are used for visualization and clustering. $\nu_{gn}$ provides a batch-corrected, size-factor normalized estimate of gene expression for each gene $g$ in each cell $n$. An added advantage of the probabilistic representation by scVI is that it provides a natural probabilistic treatment of the subsequent differential analysis, resulting in lower variance in the adopted hypothesis tests.

*Results:* scVI was evaluated for its scalability, the performance of imputation. For scalability, ScVI was shown to be faster than most nonDL algorithms and scalable to handle twice as many cells as nonDL algorithms with a fixed memory. For imputation, ScVI, together with other ZINB-based models, performed better than methods using alternative distributions. However, it underperformed for the dataset (HEMATO) with fewer cells. For the latent space, scVI was shown to provide a comparable stratification of cells into previously annotated cell types. Although scVI failed to ravel SIMLR, it is among the best in capturing biological structures (hierarchical structure, dynamics, etc.) and recognizing noise in data. For batch correction, it outperforms ComBat. For normalizing sequencing depth, the size factor inferred by scVI was shown to be strongly correlated with the sequencing depth. Interestingly, the negative binomial distribution in the ZINB was found to explain the proportions of zero expressions in the cells, whereas the zero probability $\pi_{gn}$ is found to be more correlated with alignment errors. For differential expression analysis, scVI was shown to be among the best.

#### 4.4.2. LDVAE: linearly decoded variational autoencoder

LDVAE [81] is an adaption of scVI to improve the model interpretability but it still benefits from the scalability and efficiency of scVI. Also, this formulation applies to general VAE models and thus is not restricted to scRNA-seq analysis.

*Model.* LDVAE follows scVI's formulation but replaces the decoder $D_{\boldsymbol{\theta}_\nu}$ in Eq. (23) by a linear model

$$\boldsymbol{\nu}_n = \boldsymbol{W}\boldsymbol{z}_n, \tag{26}$$

where $\boldsymbol{W} \in \mathbb{R}^{d \times G}$ is the weight matrix. Being the linear decoder provides interpretability in the sense that the relationship between latent representation $\boldsymbol{z}_n$ and gene expression $\boldsymbol{\nu}_n$ can be readily identified. LDVAE still follows the same loss and non-linear inference scheme as scVI.

_Results._ LDVAE's latent variable $\boldsymbol{z}_n$ could be used for clustering of cells with similar accuracy as a VAE. Although LDVAE had a higher reconstruction error than VAE, due to the linear decoder, the variations along the different axes of $\boldsymbol{z}_n$ establish direct linear relationships with input genes. As an example from analyzing mouse embryo scRNA-seq, $z_{1,n}$, the second element of $\boldsymbol{z}_n$, is shown to relate to simultaneous variations in the expression of gene $Pou5f1$ and $Tdgf1$. In contrast, such interpretability would be intractable without approximation for a VAE. LDVAE was also shown to induce fewer correlations between latent variables and to improve the grouping of the regulatory programs. LDVAE is capable to scale to a large dataset with ~2M cells.

#### 4.4.3. SAUCIE

SAUCIE [15] is an AE (**Figs. 2B, 4B**) designed to perform multiple functions, including clustering, batch correlation, imputation, and visualization. SAUCIE is applied to the normalized data instead of count data.

_Model._ SAUCIE includes multiple model components designed for different functions.

1. Clustering: SAUCIE first introduced a "digital" binary encoding layer $\boldsymbol{h}^c \in \{0,1\}^J$ in the decoder $D$ that functions to encode the cluster ID. To learn this encoding, an entropy loss is introduced

$$L_D = \sum_{k=1}^{K} p_k \log p_k, \tag{27}$$

where $p_k$ is the probability (proportion) of activation on neuron $k$ by the previous layer. Minimizing this entropy loss promotes sparse neurons, thus forcing a binary encoding. To encourage clustering behavior, SAUCIE also introduced an intracluster loss as

$$L_C = \sum_{i,j:h_i^c=h_j^c} \left\|\hat{\boldsymbol{x}}_i - \hat{\boldsymbol{x}}_j\right\|^2, \tag{28}$$

which computes the distance $L_C$ between the expressions of a pair of cells ($\hat{\boldsymbol{x}}_i,\ \hat{\boldsymbol{x}}_j$) that have the same cluster ID ($h_i^c = h_j^c$).

2. Batch correction: To correct the batch effect, an MMD loss is introduced to measure the differences in terms of the distribution between batches in the latent space

$$L_B = \sum_{l=1,l\neq ref}^{B} MMD(\boldsymbol{z}_{ref}, \boldsymbol{z}_l), \tag{29}$$

where $B$ is the total number of batches and $\boldsymbol{z}_{ref}$ is the latent variable of an arbitrarily chosen reference batch.

3. Imputation and visualization: The output of the decoder is taken by SAUCIE as an imputed version of the input gene expression. To visualize the data without performing an additional dimension reduction directly, the dimension of the latent variable $\boldsymbol{z}_n$ is forced to 2.

Training the model includes two sequential runs. In the first run, an autoencoder is trained to minimize the loss $L_0 + \lambda_B L_B$ with $L_0$ being the MSE reconstruction loss defined in (9) so that a batch-corrected, imputed input $\tilde{\boldsymbol{x}}$ can be obtained at the output of the decoder. In the second run, the bottleneck layer of the encoder from the first run is replaced by a 2-D latent code for visualization and a digital encoding layer is also introduced. This model takes the cleaned $\tilde{\boldsymbol{x}}$ as the input and is trained for clustering by minimizing the loss $L_0 +$ $\lambda_D L_D + \lambda_C L_C$ . After the model is trained, $\tilde{x}$ is the imputed, batch-corrected gene expression. The 2-D latent code is used for visualization and the binary encoder encodes the cluster ID.

*Results.* SAUCIE was evaluated for clustering, batch correction, imputation, and visualization on both simulated and real scRNA-seq and scCyToF datasets. The performance was compared to minibatch *kmeans*, Phenograph [82] and 22 for clustering; MNN [32] and canonical correlation analysis (CCA) [33] for batch correction; PCA, Monocle2 [83], diffusion maps, UMAP [84], tSNE [85] and PHATE [86] for visualization; MAGIC [51], scImpute [37] and nearest neighbors completion (NN completion) for imputation. Results showed that SAUCIE had a better or comparable performance with other approaches. Also, SAUCIE has better scalability and faster runtimes than any of the other models. SAUCIE's results on the scCyToF dengue dataset were further analyzed in greater detail. SAUCIE was able to identify subtypes of the T cell populations and demonstrated distinct cell manifold between acute and healthy subjects.

#### 4.4.4. scScope:

scScope [87] is an AE (**Figs. 2B, 4D**) with recurrent steps designed for imputation and batch correction.

*Model.* scScope has the following model design for batch correction and imputation.

1. Batch correction: A batch correction layer is applied to the input expression as

$$\boldsymbol{x}_n^c = ReLu(\boldsymbol{x}_n - \boldsymbol{B}\boldsymbol{u}_c), \tag{30}$$

   where and $ReLU$ is the ReLu activation function, $\boldsymbol{B} \in \mathbb{R}^{G \times K}$ is the batch correction matrix, $\boldsymbol{u}_c \in \{0,1\}^{K \times \mathbf{1}}$ is an indicator vector with entry 1 indicates the batch of the input, and $K$ is the total number of batches.

2. Recursive imputation: Instead of using the reconstructed expression $\hat{\boldsymbol{x}}_n$as the imputed expression like in SAUCIE, scScope adds an imputer to $\hat{\boldsymbol{x}}_n$ to recursively improve the imputation result. The imputer is a single-layer autoencoder, whose decoder performs the imputation as

$$\hat{\hat{\boldsymbol{x}}}_n = P_I\big[D_I\big(\hat{\hat{\boldsymbol{z}}}_n\big)\big], \tag{31}$$

where $\hat{\hat{\boldsymbol{z}}}_n$ is the output of the imputer encoder, $D_I$ is the imputer decoder, and $P_I$ is a masking function that set the elements in $\hat{\hat{\boldsymbol{x}}}_n$ that correspond to the non-missing values to zero. Then, $\hat{\hat{\boldsymbol{x}}}_n$ will be fed back to fill the missing value in the batch corrected input as $\boldsymbol{x}_n^c + \hat{\hat{\boldsymbol{x}}}_n$, which will be passed on to the main autoencoder. This recursive imputation can iterate multiple cycles as selected.

The loss function is defined as

$$\mathcal{L}(\boldsymbol{\Theta}) = \sum\nolimits_{n=1}^{N} \sum_{t=1}^{T} ||P_I^-[\boldsymbol{x}_n^c - \hat{\boldsymbol{x}}_n^t]||^2 \tag{32}$$

where $T$ is the total number of recursion, $\hat{\boldsymbol{x}}_n^t$ is the reconstructed expression at $t$th recursion, $P_I^-$ is another masking function that forces the loss to compute only the non-missing values in $\boldsymbol{x}_n^c$.

*Results.* scScope was evaluated for its scalability, clustering, imputation, and batch correction. It was compared with PCA, MAGIC, ZINB-WaVE, SIMLR, AE, scVI, and DCA. For scalability and training speed, scScope was shown to offer scalability (for >100K cells) with high efficiency (faster than most of the approaches). For clustering results, scScope was shown to outperform most of the algorithms on small simulated datasets but offer similar performance on large simulated datasets. For batch correction, scScope performed comparably with other approaches but with faster runtime. For imputation, scScope produced smaller errors consistently across a different range of expression. scScope was further shown to be able to identify rear cell populations from a large mix of cells.

### 4.5. Automated cell type identification

scRNA-seq is able to catalog cell types in complex tissues under different conditions. However, the commonly adopted manual cell typing approach based on known markers is time-consuming and less reproducible. We survey deep learning models of automated cell type identification.

#### 4.5.1. *DigitalDLSorter*

DigitalDLSorter [88] was proposed to identify and quantify the immune cells infiltrated in tumors captured in bulk RNA-seq, utilizing single-cell RNA-seq data.

*Model.* DigitalDLSorter is a 4-layer DNN (**Fig. 4A**) (2 hidden layers of 200 neurons each and an output of 10 cell types). The DigitalDLSorter is trained with two single-cell datasets: breast cancers [89] and colorectal cancers [90]. For each cell, it is determined to be tumor cell or non-tumor cell using RNA-seq based CNV method [89], followed by using xCell algorithm [91] to determine immune cell types for non-tumor cells. Different pseudo bulk (from 100 cells) RNA-seq datasets were prepared with known mixture proportions to train the DNN. The output of DigitalDLSorter is the predicted proportions of cell types in the input bulk sample.

*Result.* DigitalDLSorter was first tested on simulated bulk RNA-seq samples. DigitalDLSorter achieved excellent agreement (linear correlation of 0.99 for colorectal cancer, and good agreement in quadratic relationship for breast cancer) at predicting cell types proportion. The proportion of immune and nonimmune cell subtypes of test bulk TCGA samples was predicted by DigitalDLSorter and the results showed a very good correlation to other deconvolution tools including TIMER [89], ESTIMATE [92], EPIC [93] and MCPCounter [94]. Using DigitalDLSorter predicted CD8+ (good prognosis for overall and disease-free survival) and Monocytes-Macrophages (MM, indicator for protumoral activity) proportions, it is found that patients with higher CD8+/MM ratio had better survival for both cancer types than those with lower CD8+/MM ratio. Both EPIC and MCPCounter yielded non-significant survival associations using their cell proportion estimate.

#### 4.5.2. scCapsNet

scCapsNet [95] is an interpretable capsule network designed for cell type prediction. The paper showed that the trained network could be interpreted to inform marker genes and regulatory modules of cell types.

_Model._ The two-layer architecture of scCapsNet takes log-transformed, normalized expressions as input to form a feature extraction network (consists of $L$ parallel single-layer neural networks) followed by a capsule network for cell-type classification (type capsules). For each L parallel feature extraction layer, it generates a primary capsule $\boldsymbol{u}_l \in \mathbb{R}^{d_p}$ as

$$\boldsymbol{u}_l = ReLU(\boldsymbol{W}_{P,l}\boldsymbol{x}_n)\ \forall l = 1, \dots, L \tag{33}$$

where $\boldsymbol{W}_{P,l} \in \mathbb{R}^{d_p \times G}$ is the weight matrix. Then, the primary capsules are fed into the capsule network to compute $K$ type capsules $\boldsymbol{v}_k \in \mathbb{R}^{d_t}$, one for each cell type, as

$$\boldsymbol{v}_k = squash\left(\sum_{l}^{L} c_{kl}\boldsymbol{W}_{kl}\boldsymbol{u}_l\right)\ \forall k = 1, \dots, K \tag{34}$$

where $squash$ is the squashing function [96] to normalize the magnitude of its input vector to be less than one, $\boldsymbol{W}_{kl}$ is another trainable weight matrix, and $c_{kl}\ \forall\ l = 1, \ldots, L$ are the coupling coefficients that represent the probability distribution of each primary capsule's impact on the prediction of cell type $k$. $c_{kl}$ is not trained but computed through the dynamic routing process proposed in the original capsule networks [95]. The magnitude of each type of capsule $\boldsymbol{v}_k$ represents the probability of a single cell $\boldsymbol{x}_n$ belonging to cell type $k$, which will be used for cell-type classification.

The training of the network minimizes the cross-entropy loss by the back-propagation algorithm. Once trained, the interpretation of marker genes and regulatory modules can be achieved by determining first the important primary capsules for each cell type and then the most significant genes for each important primary capsule (identified based on $c_{kl}$ directly). To determine the genes that are important for an important primary capsule $l$, genes are ranked base on the scores of the first principal component computed from the columns of $\boldsymbol{W}_{P,l}$ and then the markers are obtained by a greedy search along with the ranked list for the best classification performance.

*Results.* scCapsNet's performance was evaluated on human PBMCs [97] and mouse retinal bipolar cells [98] datasets and shown to have comparable accuracies (99% and 97% respectively) with DNN and other popular ML algorithms (SVM, random forest, LDA and nearest neighbor). However, the interpretability of scCapsNet was demonstrated extensively. First, examining the coupling coefficients for each cell type showed that only a few primary capsules have high values and thus are effective. Subsequently, a set of core genes were identified for each effective capsule using the greedy search on the PC-score ranked gene list. GO enrichment analysis showed that these core genes were enriched in cell-type-related biological functions. Mapping the expression data into space spanned by PCs of the columns of $\boldsymbol{W}_{P,l}$ corresponding to all core genes uncovered regulatory modules that would be missed by the T-SNE of gene expressions, which demonstrated the effectiveness of the embeddings learned by scCapsNet in capturing the functionally important features.

#### 4.5.3. netAE: network-enhanced autoencoder

netAE [99] is a VAE-based semi-supervised cell type prediction model (**Figs. 2A, 4C**) that deals with scenarios of having a small number of labeled cells.

<u>*Model.*</u> netAE works with UMI counts and assumes a ZINB distribution for $x_{gn}$ as in Eq. (22) in scVI. However, netAE adopts the general VAE loss as in Eq. (6) with two function-specific loss as

$$L(\boldsymbol{\Theta}) = -\mathcal{L}(\boldsymbol{\Theta}) + \lambda_1 \sum_{n \in S} Q(\boldsymbol{z}_n) + \lambda_2 \sum_{n \in S_L} log f(y_n | \boldsymbol{z}_n), \quad (35)$$

where $S$ is a set of indices for all cells and $S_L$ is a subset of $S$ for only cells with cell type labels, $Q$ is modified Newman and Girvan modularity [100] that quantifies cluster strength using $\boldsymbol{z}_n$, $f$ is the softmax function, and $y_n$ is the cell type label. The second loss in Eq. (34) functions as a clustering constraint and the last term is the cross-entropy loss that constrains the cell type classification.

<u>*Results:*</u> netAE was compared with popular dimension reduction methods including scVI, ZIFA, PCA and AE as well as a semi-supervised method scANVI [101]. For different dimension reduction methods, cell type classification from latent features of cells was carried out using KNN and logistic regression. The effect of different labeled samples sizes on classification performance was also investigated, where the sample size varied from as few as 10 cells to 70% of all cells. Among 3 test datasets (mouse brain cortex, human embryo development, and mouse hematopoietic stem and progenitor cells), netAE outperformed most of the baseline methods. Latent features were visualized using t-SNE and cell clusters by netAE were tighter than those by other embedding spaces. netAE also showed consistency of better cell-type classification with improved cell type clustering. This suggested that the latent spaces learned with added modularity constraint in the loss helped identify clusters of similar cells. Ablation study by removing each of the three loss terms in Eq. (35) showed a drop of cell-type classification accuracy, suggesting all three were necessary for the optimal performance.

#### 4.5.4. scDGN - supervised adversarial alignment of single-cell RNA-seq data

scDGN [102], or Single Cell Domain Generalization Network (**Fig. 4G)**, is an domain adversarial network that aims to accurately assign cell types of single cells while performing batch removal (domain adaptation) at the same time. It benefits from the superior ability of domain adversarial learning to learn representations that are invariant to technical confounders.

_Model._ scDGN takes the log-transformed, normalized expression as the input and has three main modules: i) an encoder ($E_{\boldsymbol{\phi}}(\boldsymbol{x}_n)$) for dimension reduction of scRNA-seq data, ii) cell-type classifier, $C_{\boldsymbol{\phi}_C}\left(E_{\boldsymbol{\phi}}(\boldsymbol{x}_n)\right)$ with parameters $\boldsymbol{\phi}_C$ , and iii) domain (batch) discriminator, $D_{\boldsymbol{\phi}_{\mathrm{D}}}\left(E_{\boldsymbol{\phi}}(\boldsymbol{x}_n)\right)$. The model has a Siamese design and the training takes a pair of cells $(\boldsymbol{x}_1, \boldsymbol{x}_2)$, each from the same or different batches. The encoder network contains two hidden layers with 1146 and 100 neurons. $C_{\boldsymbol{\phi}_C}$ classifies the cell type and $D_{\boldsymbol{\phi}_{\mathrm{D}}}$ predicts whether $\boldsymbol{x}_1$ and $\boldsymbol{x}_2$ are from the same batch or not. The overall loss is denoted by

$$L(\boldsymbol{\phi}, \boldsymbol{\phi}_C, \boldsymbol{\phi}_{\mathrm{D}}) = L_C\left(C_{\boldsymbol{\phi}_C}\left(E_{\boldsymbol{\phi}}(\boldsymbol{x}_1)\right)\right) - \lambda L_D\left(D_{\boldsymbol{\phi}_{\mathrm{D}}}\left(E_{\boldsymbol{\phi}}(\boldsymbol{x}_1)\right), D_{\boldsymbol{\phi}_{\mathrm{D}}}\left(E_{\boldsymbol{\phi}}(\boldsymbol{x}_2)\right)\right), \quad (36)$$

where $L_C$ is the cross-entropy loss, $L_D$ is a contrastive loss as described in [103]. Notice that (46) has an adversarial formulation and minimizing this loss maximizes the misclassification of cells from different batches, thus making them indistinguishable. Similar to GAN training, scDGN is trained to iteratively solve: $\widehat{\boldsymbol{\phi}}_D = \mathrm{argmin}_{\boldsymbol{\phi}_D} L(\widehat{\boldsymbol{\phi}}, \widehat{\boldsymbol{\phi}}_C, \boldsymbol{\phi}_{\mathrm{D}})$ and $(\widehat{\boldsymbol{\phi}}, \widehat{\boldsymbol{\phi}}_C) = \mathrm{argmin}_{\boldsymbol{\phi}, \boldsymbol{\phi}_C} L(\boldsymbol{\phi}, \boldsymbol{\phi}_C, \widehat{\boldsymbol{\phi}}_D)$.

*Results.* scDGN was tested for classifying cell types and aligning batches ranging in size from 10 to 39 cell types and from 4 to 155 batches. The performance was compared to a series of deep learning and traditional machine learning methods, including Lin et al. DNN [66], CaSTLe [104], MNN [32], scVI [17], and Seurat [10]. scDGN outperformed all other methods in the classification accuracy on a subset of scQuery datasets (0.29), PBMC (0.87), and 4 of the six Seurat pancreatic datasets (0.86-0.95). PCA visualization of the learned data representations demonstrated that scDGN overcame the batch differences and clearly separated cell clusters based on cell types, while other methods were vulnerable to batch effects. In summary, scDGN is a supervised adversarial alignment method to eliminate the batch effect of scRNA-seq data and create cleaner representations of cell types.

### 4.6. Biological function prediction

Predicting biological function and response to treatment at single cell level or cell types is critical to understand cellular system functioning and potent responses to stimulations. DL models are capable of capture gene-gene relationship and their property in the latent space. Several models we reviewed below provide exciting approaches to learn complex biological functions and outcomes.

#### 4.6.1. CNNC: convolutional neural network for coexpression

CNNC [105] is proposed to infer causal interactions between genes from scRNA-seq data.

_Model._ CNNC is a Convolutional Neural Network (CNN) (**Fig. 4F**), the most popular DL model. CNNC takes expression levels of two genes from many cells and transforms them into a 32 x 32 image-like normalized empirical probability function (NEPDF), which measures the probabilities of observing different coexpression levels between the two genes. CNNC includes 6 convolutional layers, 3 max-pooling layers, 1 flatten layer, and one output layer. All convolution layers have 32 kernels of size 3x3. Depending on the application, the output layer can be designed to predict the state of interaction (Yes/No) between the genes or the causal interaction between the input genes (no interaction, gene A regulates gene B, or gene B regulates gene A).

_Result._ CNNC was trained to predict transcription factor (TF)-Gene interactions using the mESC data from scQuery [106], where the ground truth interactions were obtained from the ChIP-seq dataset from the GTRD database [107]. The performance was compared with DNN, count statistics [108], and mutual information-based approach [109]. CNNC was shown to have more than 20% higher AUPRC than other methods and reported almost no false-negative for the top 5% predictions. CNNC was also trained to predict the pathway regulator-target gene pairs. The positive regulator-gene pairs were obtained from KEGG [110], Reactome [111], and negative samples were genes pairs that appeared in pathways but not interacted. CNNC was shown to have better performance of predicting regulator-gene pairs on both KEGG and Reactome pathways than other methods including Pearson correlation, count statistics, GENIE3 [112], Mutual information, Bayesian directed network (BDN), and DREMI [109]. CNNC was also applied for causality prediction between two genes, that is if two genes regulate each other and if they do, which gene is the regulator. The ground truth causal relationships were also obtained from the KEGG and Reactome datasets. Again, CNNC reported better performance than BDN, the common method developed to learn casual relationships from gene expression data. CNNC was finally trained to assign 3 essential cell functions (cell cycle, circadian rhythm, and immune system) to genes. This is achieved by training CNNC to predict pairs of genes from the same function (e.g. Cell Cycle defined by mSigDB from gene set enrichment analysis (GSEA) [113]) as 1 and all other pairs as 0. The performance was compared with "guilt by association" and DNN, and CNNC was shown to have more than 4% higher AUROC and reported all positives for the top 10% predictions.

#### 4.6.2. scGen, a generative model to predict perturbation response of single cells across cell types

scGen [114] is designed to learn cell response to certain perturbation (drug treatment, gene knockout, etc) from single cell expression data and predict the response to the same perturbation for a new sample or a new cell type. The novelty of scGen is that it learns the response in the latent space instead of the expression data space.

*Model.* ScGen follows the general VAE (**Figs. 2A, 4C**) for scRNA-seq data but uses the "latent space arithmetics" to learn perturbations' response. Given scRNA-seq samples of perturbed (denoted as *p*) and unperturbed cells (denoted as *unp*), a VAE model is trained.

Then, the latent space representation $\boldsymbol{z}_p$ and $\boldsymbol{z}_{unp}$ are obtained for the perturbed and unperturbed cells. Following the notion that VAE could map nonlinear operations (e.g., perturbation) in the data space to linear operations in the latent space, ScGen estimate the response in the latent space as $\boldsymbol{\delta} = \bar{\boldsymbol{z}}_p - \bar{\boldsymbol{z}}_{unp}$, where $\bar{\boldsymbol{z}}_{\cdot}$ is the average representation of samples from the same cell type or different cell types. Then, given the latent representation $\boldsymbol{z'}_{unp}$ of an unperturbed cell for a new sample from the same or different cell type, the latent representation of the corresponding perturbed cell can be predicted as $\boldsymbol{z'}_p$=$\boldsymbol{z'}_{unp} + \boldsymbol{\delta}$. The expression of the perturbed cell can also be estimated by feeding $\boldsymbol{z'}_p$ into the VAE decoder. The scGen can also be expanded to samples and treatment across two species (using orthologues between species). When scGen is trained for species 1, or $s_1$, with both perturbed and unperturbed cells but species 2, $s_2$, with only unperturbed cells, then the latent code for the perturbed cells from $s_2$ can be predicted as $\boldsymbol{z}_{s_2,p} = \frac{1}{2}\left(z_{s_1,p} + z_{s_2,unp} + \delta_s + \delta_p\right)$ where $\boldsymbol{\delta}_p = z_{s_1,unp} - z_{s_1,p}$ captures the response of perturbation and $\boldsymbol{\delta}_s = z_{s_1} - z_{s_2}$ represents the difference between species.

*Result.* scGen was applied to predict perturbation of out-of-samples response in human PBMCs data, and scGen showed a higher average correlation ($R^2$= 0.948) between predicted and real data for six cell types. Comparing with other methods including CVAE [115], style transfer GAN [116], linear approaches based on vector arithmetics(VA) [114] and PCA+VA, scGen predicted full distribution of ISG15 gene (strongest regulated gene by IFN-β) response to IFN- β [117], while others might predict mean (CAVE and style transfer GAN) but failed to produce the full distribution. scGen was also tested on predicting the intestinal epithelial cells' response to infection[118]. For early transit-amplifying cells, scGen showed good prediction ($R^2$=0.98) for both H. poly and Salmonella infection. Finally, scGen was evaluated for perturbation across species using scRNA-seq data set by Hagai et al [119], which comprises bone marrow-derived mononuclear phagocytes from mice, rats, rabbits, and pigs perturbed with lipopolysaccharide (LPS). scGen's predictions of LPS perturbation responses were shown to be highly correlated ($R^2$ = 0.91) with the real responses.

## 5. Conclusions

We systematically survey 25 DL models according to the challenges they address. Unlike other surveys, we categorize major DL model statements into VAE, AE, and GAN with a unified mathematic formulation in Section 3 (graphic model representation in **Fig. 2**), which will guide readers to focus on the DL model selection, training strategies, and loss functions in each algorithm. Specifically, the differences in loss functions are highlighted in each DL model's applications to meet specific objectives. DL/ML models that 25 surveyed models are evaluated against are presented in **Fig. 3**, providing a straightforward way for readers to pick up the most suitable DL model at a specific step for their own scRNA-seq data analysis. All evaluation methods are listed in **Table 3**, as as we foresee to be an easy recipe book for researchers to establish their scRNA-seq pipeline. In addition, a summary of all the 25 DL models concerning their DL models, evaluation metrics, implementation environment, and downloadable source codes is presented in **Table 1**. Taking together, this survey will provide a rich resource for DL method selection for proper research applications, and we expect to inspire new DL model development for scRNA-seq analysis.

In this review, we focus our survey on common DL models, such as AE, VAE, and GAN, and their model variations or combinations to address single-cell data analysis challenges. With the advancement of multi-omics single-cell technologies, such as cyTOF (such as SAUCIE [15]), spatial transcriptome using DNN [120], and CITE-seq that simultaneously generates read counts for surface protein expression along with gene expression [121, 122]. Other than 3 most common DL models, we also include network frameworks such as Capsule NN (such as scCapsNet [95]), Convolution NN (such as CNNC [105]) and domain adaption learning (such as scDGN [102]). It is expected that more DL models will be developed and implemented for the most challenging step for scRNA-seq data, including but not limited to, data interpretation. For example, integrating protein-protein interaction graphs into DL models has been shown for its advantages of biological knowledge and nonlinear interactions embedded in the graphs [123-125]. Indeed, a recently published scRNA-seq analysis pipeline, scGNN [126], incorporates 3 iterative autoencoders (including one graph autoencoder) and successfully demonstrated Alzheimer's disease-related neural development and differentiation mechanism. We expect that our careful organization of this review paper will provide a basic understanding of DL models for scRNA-seq and a list of critical elements to be considered for future developments in DLs.

**Funding**

This article's publication costs were supported partially by the National Institutes of Health (CTSA 1UL1RR025767-01 to YC, R01GM113245 to YH, NCI Cancer Center Shared Resources P30CA54174 to YC, and K99CA248944 to YCC); National Science

Foundation (2051113 to YFJ); Cancer Prevention and Research Institute of Texas (RP160732 to YC, RP190346 to YC and YH); and the Fund for Innovation in Cancer Informatics (ICI Fund to YCC and YC). The funding sources had no role in the design of the study; collection, analysis, and interpretation of data; or in writing the manuscript.

## Authors' contributions

YH, YC, MF and YFJ conceived the study. MF, ZL, TZ, MMH, YCC, ZY, KP, SJ, JZ, SJG, YFJ, YC and YH summarized resources, wrote, and approved the final version of paper.

## Figure Captions

**Figure 1. Single cell data analysis steps for both conventional ML methods (bottom) and DL methods (top).** Depending on the input data and analysis objectives, major scRNA-seq analysis steps are illustrated in the the center flow chart (color boxes) with convential ML approaches along with optional analysis modules below each analysis step. Deep learning approaches are categorized as Deep Neural Network, Generalive Adversarial Network, Variational Autoencoder, and Autoencoder. For each DL approach, optional algorithms are listed on top of each step.

**Figure 2. Graphical models of the major surveyed DL models including A)** Variational Autoencoder **B)** Autoencoder; and **C)** Generative Adversarial Network

**Figure 3**. **Algorithm comparison grid.** DL methods surveyed in the paper are listed in the left-hand side, and some in the column. Algorithms selected to compare in each DL method are marked by "■" at each cross-point.

**Figure 4**. **DL model network illustration**. **A**) Deep neural network, **B**) Autoencoder, **C**) Variational autoencoder, **D**) Autoencoder with recursive imputer, **E**) Generative adversarial network, **F**) Convolutional neural network, and **G**) Domain adversarial neural network.

# Tables

**Table 1.** Deep Learning algorithms reviewed in the paper

| App | Algorithm | Models | Evaluation | Environment | Codes | Refs |
|---|---|---|---|---|---|---|
| **Imputation** | | | | | | |
| | DCA | AE | DREMI | Keras, Tensorflow, scanpy | https://github.com/theislab/dca | [18] |
| | SAVER-X | AE+TL | t-SNE, ARI | R/sctransfer | https://github.com/jingshuw/SAVERX | [52] |
| | DeepImpute | DNN | MSE, Pearson's correlation | Keras/Tensorflow | https://github.com/lanagarmire/DeepImpute | [20] |
| | LATE | AE | MSE | Tensorflow | https://github.com/audreyqyfu/LATE | [53] |
| | scGAMI | AE | NMI, ARI, HS and CS | Tensorflow | https://github.com/QUST-AIBBDRC/scGMAI/ | [54] |
| | scIGANs | GAN | ARI, ACC, AUC, and F-score | PyTorch | https://github.com/xuyungang/scIGANs | [19] |
| **Batch correction** | | | | | | |
| | BERMUDA | AE+TL | kBET, entropy of Mixing, SI | PyTorch | https://github.com/txWang/BERMUDA | [58] |
| | DESC | AE | ARI, KL | Tensorflow | https://github.com/eleozzr/desc | [62] |
| | iMAP | AE+GAN | kBET, LISI | PyTorch | https://github.com/Svvord/iMAP | [65] |
| **Clustering, latent representation, dimension reduction, and data augmentation** | | | | | | |
| | Dhaka | VAE | ARI, Spearman Correlation | Keras/Tensorflow | https://github.com/MicrosoftGenomics/Dhaka | [67] |
| | scvis | VAE | KNN preservation, log-likelihood | Tensorflow | https://bitbucket.org/jerry00/scvis-dev/src/master/ | [70] |
| | scVAE | VAE | ARI | Tensorflow | https://github.com/scvae/scvae | [71] |
| | VASC | VAE | NMI, ARI, HS, and CS | H5py, keras | https://github.com/wang-research/VASC | [72] |
| | scDeepCluster | AE | ARI, NMI, clustering accuracy | Keras, Scanpy | https://github.com/ttgump/scDeepCluster | [74] |

| | | | | | |
|---|---|---|---|---|---|
| cscGAN | GAN | t-SNE, marker genes, MMD, AUC | Scipy, Tensorflow | https://github.com/imsb-uke/scGAN | [77] |
| **Multi-functional models** (IM: imputation, BC: batch correction, CL: clustering) | | | | | |
| scVI | VAE | **IM**: $L_1$ distance; **CL**: ARI, NMI, SI; **BC**: Entropy of Mixing | PyTorch, Anndata | https://github.com/YosefLab/scvi-tools | [17] |
| LDVAE | VAE | Reconstruction errors | Part of scVI | https://github.com/YosefLab/scvi-tools | [81] |
| SAUCIE | AE | **IM**: $R^2$ statistics; **CL**: SI; **BC**: modified kBET; Visualization: Precision/Recall | Tensorflow | https://github.com/KrishnaswamyLab/SAUCIE/ | [15] |
| scScope | AE | **IM**:Reconstruction errors; **BC**: Entropy of mixing; **CL**: ARI | Tensorflow, Scikit-learn | https://github.com/AltschulerWu-Lab/scScope | [87] |
| **Cell type Identification** | | | | | |
| DigitalDLSorter | DNN | Pearson correlation | R/Python/Keras | https://github.com/cartof/digitalDLSorter | [88] |
| scCapsNet | CapsNet | Cell-type Prediction accuracy | Keras, Tensorflow | https://github.com/wanglf19/scCaps | [95] |
| netAE | VAE | Cell-type Prediction accuracy, t-SNE for visualization | pyTorch | https://github.com/LeoZDong/netAE | [99] |
| scDGN | DANN | Prediciton accuracy | pyTorch | https://github.com/SongweiGe/scDGN | [102] |
| **Function analysis** | | | | | |
| CNNC | CNN | AUROC, AUPRC, and accuracy | Keras, Tensorflow | https://github.com/xiaoyeye/CNNC | [105] |
| scGen | VAE | Correlation, visualization | Tensorflow | https://github.com/theislab/scgen | [114] |

DL Model keywords: AE: autoencoder, AE+TL: autoencoder with transfer learning, AE: variational autoencoder, GAN: Generative adversarial network, CNN: convolutional neural network, DNN: deep neural network, DANN: domain adversarial neural network, CapsNet: capsule neural network

**Table 2a:** Simulated single-cell data/algorithms

| Title | Algorithm | # Cells | Simulation methods | Reference |
|---|---|---|---|---|
| **Splatter** | DCA, DeepImpute, PERMUDA, scDeepCluster, scVI, scScope, solo | ~2000 | Splatter/R | [79] |
| **CIDR** | scIGAN | 50 | CIDR simulation | [55] |
| **NB+dropout** | Dhaka | 500 | Hierachical model of NB/Gamma + random dropout | |
| **Bulk RNA-seq** | SAUCIE | 1076 | 1076 CCLE bulk RNAseq + dropout conditional on expression level | |
| **SIMLR** | scScope | 1 million | SIMLR, high-dimensional data generated from latent vector | [44] |

**Table 2b:** Human single-cell data sources used by different DL algorithms

| Title | Algorithm | Cell origin | # Cells | Data Sources | Reference |
|---|---|---|---|---|---|
| **68k PBMCs** | DCA<br>SAVER-X<br>LATE, scVAE, scDeepCluster, scCapsNet, scDGN | Blood | 68,579 | 10X Single Cell Gene Expression Datasets | |
| **Human pluripotent** | DCA | hESCs | 1,876 | GSE102176 | [127] |
| **CITE-seq** | SAVER-X | Cord blood mononuclear cells | 8,005 | GSE100866 | [128] |
| **Midbrain and Dopaminergic Neuron Development** | SAVER-X | Brain/ embryo ventral midbrain cells | 1,977 | GSE76381 | [129] |
| **HCA** | SAVER-X | Immune cell, Human Cell Atlas | 500,000 | HCA data portal | |
| **Breast tumor** | SAVER-X | Immune cell in tumor micro-environment | 45,000 | GSE114725 | [130] |
| **293T cells** | DeepImpute, iMAP | Embryonic kidney | 13,480 | 10X Single Cell Gene Expression Datasets | |
| **Jurkat** | DeepImpute, iMAP | Blood/ lymphocyte | 3,200 | 10X Single Cell Gene Expression Datasets | |
| **ESC, Time-course** | scGAN | ESC | 350, 758 | GSE75748 | [131] |
| **Baron-Hum-1** | scGMAI, VASC | Pancreatic islets | 1,937 | GSM2230757 | [132] |
| **Baron-Hum-2** | scGMAI, VASC | Pancreatic islets | 1,724 | GSM2230758 | [132] |
| **Camp** | scGMAI, VASC | Liver cells | 303 | GSE96981 | [133] |
| **CEL-seq2** | PERMUDA, DESC | Pancreas/Islets of Langerhans | | GSE85241 | [134] |
| **Darmanis** | scGMAI, scIGAN, VASC | Brain/cortex | 466 | GSE67835 | [135] |
| **Tirosh-brain** | Dhaka, scvis | Oligodendroglioma | >4800 | GSE70630 | [136] |
| **Patel** | Dhaka | Primary glioblastoma cells | 875 | GSE57872 | [137] |
| **Li** | scGMAI, VASC | Blood | 561 | GSE146974 | [62] |
| **Tirosh-skin** | scvis | melanoma | 4645 | GSE72056 | [68] |
| **xenograft 3, and 4** | Dhaka | Breast tumor | ~250 | EGAS00001002170 | [138] |
| **Petropoulos** | VASC/netAE | Human embryos | 1,529 | E-MTAB-3929 | |
| **Pollen** | scGMAI, VASC | | 348 | SRP041736 | [139] |
| **Xin** | scGMAI, VASC | Pancreatic cells (α-, β-, δ-) | 1,600 | GSE81608 | [140] |
| **Yan** | scGMAI, VASC | embryonic stem cells | 124 | GSE36552 | [141] |
| **PBMC3k** | VASC, scVI | Blood | 2,700 | SRP073767 | [97] |
| **CyTOF, Dengue** | SAUCIE | Dengue infection | 11 M, ~42 antibodies | Cytobank, 82023 | [15] |
| **CyTOF, ccRCC** | SAUCIE | Immunue profile of 73 ccRCC patients | 3.5 M, ~40 antibodies | Cytobank: 875 | [142] |
| **CyTOF, breast** | SAUCIE | 3 patients | | Flow Repository: FR-FCM-ZYJP | [130] |
| **Chung, BC** | DigitalDLSorter | Breast tumor | 515 | GSE75688 | [89] |
| **Li, CRC** | DigitalDLSorter | Colorectal cancer | 2,591 | GSE81861 | [90] |
| **Pancreatic datasets** | scDGN | Pancreas | 14693 | SeuratData | |
| **Kang, PBMC** | scGen | PBMC stimulated by INF-β | ~15,000 | GSE96583 | [117] |

**Table 2c:** Mouse single-cell data sources used by different DL algorithms

| Title | Algorithm | Cell origin | # Cells | Data Sources | Reference |
|---|---|---|---|---|---|
| **Brain cells from E18 mice** | DCA, SAUCIE | Brain Cortex | 1,306,127 | 10x: Single Cell Gene Expression Datasets | |
| **Midbrain and Dopaminergic Neuron Development** | SAVER-X | Ventral Midbrain | 1907 | GSE76381 | [129] |
| **Mouse cell atlas** | SAVER-X | | 405,796 | GSE108097 | [143] |
| **neuron9k** | DeepImpute | Cortex | 9128 | 10x: Single Cell Gene Expression Datasets | |
| **Mouse Visual Cortex** | DeepImpute | Brain cortex | 114601 | GSE102827 | [144] |
| **murine epidermis** | DeepImpute | Epidermis | 1422 | GSE67602 | [145] |
| **myeloid progenitors** | LATE<br>DESC, SAUCIE | Bone marrow | 2,730 | GSE72857 | [146] |
| **Cell-cycle** | scIGAN | mESC | 288 | E-MTAB-2805 | [147] |
| **A single-cell survey** | | Intestine | 7721 | GSE92332 | [118] |
| **Tabula Muris** | iMAP | Mouse cells | >100K | | |
| **Baron-Mou-1** | VASC | Pancreas | 822 | GSM2230761 | [132] |
| **Biase** | scGMAI, VASC | Embryos/SMARTer | 56 | GSE57249 | [148] |
| **Biase** | scGMAI, VASC | Embryos/Fluidigm | 90 | GSE59892 | [148] |
| **Deng** | scGMAI, VASC | Liver | 317 | GSE45719 | [149] |
| **Klein** | VASC<br>scDeepCluster<br>scIGAN | Stem Cells | 2,717 | GSE65525 | [150] |
| **Goolam** | VASC | Mouse Embryo | 124 | E-METAB-3321 | [151] |
| **Kolodziejczyk** | VASC | mESC | 704 | E-MTAB-2600 | [152] |
| **Usoskin** | VASC | Lumbar | 864 | GSE59739 | [153] |
| **Zeisel** | VASC, scVI, SAUCIE, netAE | Cortex, hippocampus | 3,005 | GSE60361 | [154] |
| **Bladder cells** | scDeepCluster | Bladder | 12,884 | GSE129845 | [155] |
| **HEMATO** | scVI | Blood cell | >10,000 | GSE89754 | [156] |
| **retinal bipolar cells** | scVI, scCapsNet<br>SAUCIE | retinal | ~25,000 | GSE81905 | [98] |
| **Embryo at 9 time points** | LDAVE | embryos from E6.5 to E8.5 | 116,312 | GSE87038 | [157] |
| **Embryo at 9 time points** | LDAVE | embryos from E9.5 to E13.5 | ~2 millions | GSE119945 | [158] |
| **CyTOF,** | SAUCIE | Mouse thymus | 200K, ~38 antibodies | Cytobank: 52942 | [159] |
| **Nestorowa** | netAE | hematopoietic stem and progenitor cells | 1,920 | GSE81682 | [160] |
| **small intestinal epithelium** | scGen | Infected with Salmonella and worm H. polygyrus | 1,957 | GSE92332 | [118] |

**Table 2d:** Single-cell data derived from other species

| Title | Algorithm | Species | Tissue | # Cells | SRA/GEO | Reference |
|---|---|---|---|---|---|---|
| **Worm neuron cells**[1] | scDeepCluster | C. elegans | Neuron | 4,186 | GSE98561 | [161] |
| **Cross species, stimulation with LPS and dsRNA** | scGen | Mouse, rat, rabbit, and pig | bone marrow-derived phagocyte | 5,000 to 10,000 /species | 13 accessions in ArrayExpress | [119] |

1. Processed data is available at https://github.com/ttgump/scDeepCluster/tree/master/scRNA-seq%20data

**Table 2e:** Large single-cell data source used by various algorithms

| Title | Sources | Notes |
|---|---|---|
| **10X Single-cell gene expression dataset** | https://support.10xgenomics.com/single-cell-gene-expression/datasets | Contains large collection of scRNA-seq dataset generated using 10X system |
| **Tabula Muris** | https://tabula-muris.ds.czbiohub.org/ | Compendium of scRNA-seq data from mouse |
| **HCA** | https://data.humancellatlas.org/ | Human single-cell atlas |
| **MCA** | https://figshare.com/s/865e694ad06d5857db4b, or GSE108097 | Mouse single-cell atlas |
| **scQuery** | https://scquery.cs.cmu.edu/ | A web server cell type matching and key gene visualization. It is also a source for scRNA-seq collection (processed with common pipeline) |
| **SeuratData** | https://github.com/satijalab/seurat-data | List of datasets, including PBMC and human pancreatic islet cells |
| **cytoBank** | https://cytobank.org/ | Community of big data cytometry |

**Table 3**. Evaluation metrics used in surveyed DL algorithms

| Evaluation Method | Equations | Explanation |
|---|---|---|
| Pseudobulk RNA-seq | | Average of normalized (log2-transformed) scRNA-seq counts across cells is calculated and then correlation coefficient between the pseudobulk and the actual bulk RNA-seq profile of the same cell type is evaluated. |
| Mean squared error (MSE) | $MSE = \frac{1}{n}\sum_{i=1}^{n}(x_i - \hat{x}_i)^2$ | MSE assesses the quality of a predictor, or an estimator, from a collection of observed data $x$, with $\hat{x}$ being the predicted values. |
| Pearson correlation | $\rho_{X,Y} = \frac{cov(X,Y)}{\sigma_X \sigma_Y}$ | where cov() is the covariance, $\sigma_X$ and $\sigma_Y$ are the standard deviation of *X* and *Y*, respectively. |
| Spearman correlation | $\rho_s = \rho_{r_X,r_Y} = \frac{cov(r_X, r_Y)}{\sigma_{r_X}\sigma_{r_Y}}$ | The Spearman correlation coefficient is defined as the Pearson correlation coefficient between the rank variables, where $r_X$ is the rank of X. |
| Entropy of accuracy, $H_{acc}$ [21] | $H_{acc} = -\frac{1}{M}\sum_{i=1}^{M}\sum_{j=1}^{N_i} p_i(x_j)\log\left(p_i(x_j)\right)$ | Measures the diversity of the ground-truth labels within each predicted cluster group. $p_i(x_j)$ (or $q_i(x_j)$) are the proportions of cells in the $j^{th}$ ground-truth cluster (or predicted cluster) relative to the total number of cells in the $i^{th}$ predicted cluster (or ground-truth clusters), respectively. |
| Entropy of purity, $H_{pur}$ [21] | $H_{pur} = -\frac{1}{N}\sum_{i=1}^{N}\sum_{j=1}^{M_i} q_i(x_j)\log\left(q_i(x_j)\right)$ | Measures the diversity of the predicted cluster labels within each ground-truth group |
| Entropy of mixing [32] | $E = \sum_{i=1}^{C} p_i \log(p_i)$ | This metric evaluates the mixing of cells from different batches in the neighborhood of each cell. C is the number of batches, and $p_i$ is the proportion of cells from batch $i$ among *N* nearest cells. |
| Mutual Information (MI) [162] | $MI(U,V) = \sum_{i=1}^{\lvert U\rvert}\sum_{j=1}^{\lvert V\rvert} P_{UV}(i,j)\log\left(\frac{P_{UV}(i,j)}{P_U(i)P_V(j)}\right)$ | where $P_U(i) = \frac{\lvert U_i\rvert}{N}$ and $P_V(j) = \frac{\lvert V_j\rvert}{N}$. Also, define the joint distribution probability is $P_{UV}(i,j) = \frac{\lvert U_i \cap V_j\rvert}{N}$. The MI is a measure of mutual dependency between two cluster assignments *U* and *V*. |
| Normalized Mutual Information (NMI) [163] | $NMI(U,V) = \frac{2 \times MI(U,V)}{[H(U) + H(V)]}$ | where $H(U) = \sum P_U(i)\log\left(P_U(i)\right), H(V) = \sum P_V(i)\log\left(P_V(i)\right)$. The NMI is a normalization of the MI score between 0 and 1. |
| Kullback–Leibler (KL) divergence [164] | $D_{KL}(P\|\|Q) = \sum_{x\in\chi} P(x)\log\left(\frac{P(x)}{Q(x)}\right)$ | where discrete probability distributions *P* and *Q* are defined on the same probability space $\chi$. This relative entropy is the measure for directed divergence between two distributions. |
| Jaccard Index | $J(U,V) = \frac{\lfloor U\cap V\rfloor}{\lfloor U\cup V\rfloor}$ | 0 ≤ *J*(*U*,*V*) ≤ 1. J = 1 if clusters *U* and *V* are the same. If *U* are *V* are empty, J is defined as 1. |
| Fowlkes-Mallows Index for two clustering algorithms (FM) | $FM = \sqrt{\frac{TP}{TP+FP}\times\frac{TP}{TP+FN}}$ | TP as the number of pairs of points that are present in the same cluster in both *U* and *V*; FP as the number of pairs of points that are present in the same cluster in *U* but not in *V;* FN as the number of pairs of points that are present in the same cluster in V but not in *U;* and TN as the number of pairs of points that are in different clusters in both *U* and *V*. |
| Rand index (RI) | $RI = (a+b)/\binom{n}{2}$ | Measure of constancy between two clustering outcomes, where *a* (or *b*) is the count of number of pairs of cells in one cluster (or different clusters) from one clustering algorithm but also fall in the same cluster (or different clusters) from the other clustering algorithm. |
| Adjusted Rand index (ARI) [165] | $ARI = \frac{RI - E[RI]}{\max(RI) - E[RI]}$ | ARI is a corrected-for-chance version of *RI,* where *E*[*RI*] is the expected Rand Index. |
| Silhouette index | $s(i) = \frac{b(i) - a(i)}{\max\,(a(i), b(i))}$ | where *a*(*i*) is the average dissimilarity of $i^{th}$ cell to all other cells in the same cluster, and *b*(*i*) is the average dissimilarity of $i^{th}$ cell to all cells in the closest cluster. The range of *s*(*i*) is [−1,1], with 1 to be well-clustered and -1 to be completely misclassified. |
| Maximum Mean Discrepancy (MMD) [59] | $MMD(F,p,q) = \sup_{f\in F}\left\|\mu_p - \mu_q\right\|_f$ | MMD is a non-parametric distance between distributions based on the reproducing kernel Hilbert space, or, a distance-based measure between two distribution p and q based on the mean embeddings $\mu_p$ and $\mu_q$ in a reproducing kernel Hilbert space F. |
| k-Nearest neighbor batch-effect test (kBET) [166] | $a_n^k = \sum_{l=1}^{L} \frac{(N_{nl}^k - k\cdot f_l)^2}{k\cdot f_l} \sim X_{L-1}^2$ | Given a dataset of $N$ cells from $L$ batches with $N_l$ denoting the number of cells in batch $l$, $N_{nl}^k$ is the number of cells from batch $l$ in the *k*-nearest neighbors of cell $n$, $f_l$ is the global fraction of cells in batch $l$, or $f_l = \frac{N_l}{N}$, and $X_{L-1}^2$ denotes the $X^2$ distribution with $L-1$ degrees of freedom. It uses a $X^2$-based test for random neighborhoods of fixed size to determine the significance ("well-mixed"). |
| Local Inverse Simpson's Index (LISI) [34] | $$\frac{1}{\lambda(n)} = \frac{1}{\Sigma_{l=1}^{L}\left(p(l)\right)^2}$$ | This is the inverse Simpson's Index in the *k*-nearest neighbors of cell $n$ for all batches, where $p(l)$ denotes the proportion of batch $l$ in the *k*-nearest neighbors. The score reports the effective number of batches in the *k*-nearest neighbors of cell $n$. |
| Homogeneity | $$HS = 1 - \frac{H(P(U\|V))}{H\left(P(U)\right)}$$ | where *H*() is the entropy, and *U* is the ground-truth assignment and *V* is the predicted assignment. The *HS* range from 0 to 1, where 1 indicates perfectly homogeneous labeling. |
| Completeness | $$CS = 1 - \frac{H\left(P(V\|U)\right)}{H\left(P(V)\right)}$$ | Its values range from 0 to 1, where 1 indicates all members from a ground-truth label are assigned to a single cluster. |
| V-Measure [167] | $$V_\beta = \frac{(1+\beta)HS \times CS}{\beta HC + CS}$$ | where $\beta$ indicates the weight of *HS*. V-Measure is symmetric, *i.e.* switching the true and predicted cluster labels does not change V-Measure. |
| Precision, recall | $$Precision = \frac{TP}{TP+FP}, recall = \frac{TP}{TP+FN}$$ | TP: true positive, FP: false positive, FN, false negative. |
| Accuracy | $$Accuracy = \frac{TP+TN}{N}$$ | N: all samples tested, TN: true negative |
| $F_1$-score | $$F_1 = \frac{2\ Precision \cdot Recall}{Precision + Recall}$$ | A harmonic mean of precision and recall. It can be extended to $F_\beta$ where $\beta$ is a weight between precision and recall (similar to V-measure). |
| AUC, RUROC | 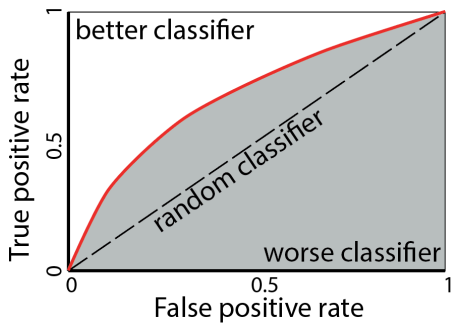 | Area Under Curve (grey area). Receiver operating characteristic (ROC) curve (red line). The similar measure can be performed on Precision-Recall curve (PRC), or AUPRC. Precision-Recall curves summarize the trade-off between the true positive rate and the positive predictive value for a predictive model (mostly for an imbalanced dataset). |